\documentclass[twoside,twocolumn,9pt]{article}
\usepackage{extsizes}
\usepackage{stfloats}
\usepackage[super,sort&compress,comma]{natbib} 
\usepackage[left=1.5cm, right=1.5cm, top=1.785cm, bottom=2.0cm]{geometry}
\usepackage{balance}
\usepackage{mathptmx}
\usepackage{sectsty}
\usepackage{graphicx}
\graphicspath{{Figs/}}
\usepackage{placeins}
\usepackage{lastpage}
\usepackage[format=plain,justification=justified,singlelinecheck=false,font={stretch=1.125,small,sf},labelfont=bf,labelsep=space]{caption}
\usepackage{float}
\usepackage{subcaption}
\usepackage{fancyhdr}
\usepackage{fnpos}
\usepackage[english]{babel}
\addto{\captionsenglish}{%
  
}
\usepackage{array}
\usepackage{amssymb}
\usepackage{booktabs}
\usepackage{charter}
\usepackage[T1]{fontenc}
\usepackage[usenames,dvipsnames]{xcolor}
\usepackage{setspace}
\usepackage[compact]{titlesec}
\usepackage{hyperref}

\definecolor{cream}{RGB}{222,217,201}

\begin{document}

\pagestyle{fancy}
\thispagestyle{plain}
\fancypagestyle{plain}{
\renewcommand{\headrulewidth}{0pt}
}

\makeFNbottom
\makeatletter
\renewcommand\LARGE{\@setfontsize\LARGE{15pt}{17}}
\renewcommand\Large{\@setfontsize\Large{12pt}{14}}
\renewcommand\large{\@setfontsize\large{10pt}{12}}
\renewcommand\footnotesize{\@setfontsize\footnotesize{7pt}{10}}
\makeatother

\renewcommand{\thefootnote}{\fnsymbol{footnote}}
\renewcommand\footnoterule{\vspace*{1pt}%
\color{cream}\hrule width 3.5in height 0.4pt \color{black}\vspace*{5pt}} 
\setcounter{secnumdepth}{5}

\makeatletter 
\renewcommand\@biblabel[1]{#1}            
\renewcommand\@makefntext[1]%
{\noindent\makebox[0pt][r]{\@thefnmark\,}#1}
\makeatother 
\renewcommand{\figurename}{\small{Fig.}~}
\sectionfont{\sffamily\Large}
\subsectionfont{\normalsize}
\subsubsectionfont{\bf}
\setstretch{1.125} %
\setlength{\skip\footins}{0.8cm}
\setlength{\footnotesep}{0.25cm}
\setlength{\jot}{10pt}
\titlespacing*{\section}{0pt}{4pt}{4pt}
\titlespacing*{\subsection}{0pt}{15pt}{1pt}

\fancyfoot{}
\fancyfoot[LO,RE]{\vspace{-7.1pt}\includegraphics[height=9pt]{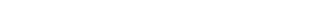}}
\fancyfoot[CO]{\vspace{-7.1pt}\hspace{13.2cm}\includegraphics{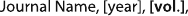}}
\fancyfoot[CE]{\vspace{-7.2pt}\hspace{-14.2cm}\includegraphics{head_foot/RF}}
\fancyfoot[RO]{\footnotesize{\sffamily{1--\pageref{LastPage} ~\textbar  \hspace{2pt}\thepage}}}
\fancyfoot[LE]{\footnotesize{\sffamily{\thepage~\textbar\hspace{3.45cm} 1--\pageref{LastPage}}}}
\fancyhead{}
\renewcommand{\headrulewidth}{0pt} 
\renewcommand{\footrulewidth}{0pt}
\setlength{\arrayrulewidth}{1pt}
\setlength{\columnsep}{6.5mm}
\setlength\bibsep{1pt}

\makeatletter 
\newlength{\figrulesep} 
\setlength{\figrulesep}{0.5\textfloatsep} 

\newcommand{\topfigrule}{\vspace*{-1pt}%
\noindent{\color{cream}\rule[-\figrulesep]{\columnwidth}{1.5pt}} }

\newcommand{\botfigrule}{\vspace*{-2pt}%
\noindent{\color{cream}\rule[\figrulesep]{\columnwidth}{1.5pt}} }

\newcommand{\dblfigrule}{\vspace*{-1pt}%
\noindent{\color{cream}\rule[-\figrulesep]{\textwidth}{1.5pt}} }

\makeatother

\twocolumn[
  \begin{@twocolumnfalse}
{\includegraphics[height=30pt]{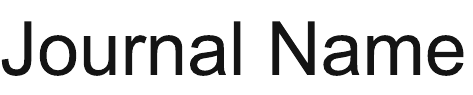}\hfill\raisebox{0pt}[0pt][0pt]{\includegraphics[height=55pt]{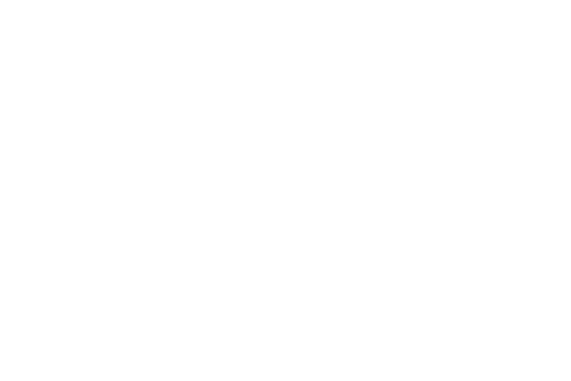}}\\[1ex]
\includegraphics[width=18.5cm]{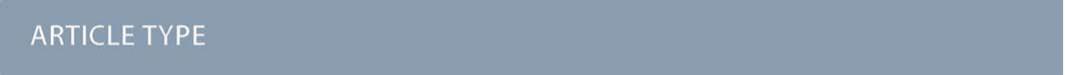}}\par
\vspace{1em}
\sffamily
\begin{tabular}{m{4.5cm} p{13.5cm} }

\includegraphics{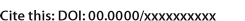} & \noindent\LARGE{\textbf{Prussian blue and Prussian white thin-film electrodes:
Single-step electrodeposition for anode and cathode applications$^\dag$}} \\%
\vspace{0.3cm} & \vspace{0.3cm} \\

 & \noindent\large{Michael Pohlitz\textit{$^{a}$}, Mohammed Fayis Kalady\textit{$^{b}$}, Larissa de O. Garcia\textit{$^{a}$}, Lindiomar Borges de Avila Jr\textit{$^{c}$}, Darius Pohl\textit{$^{d}$}, Falk Röder\textit{$^{b}$}, Flavio Abreu Araujo\textit{$^{c}$}, Axel Lubk\textit{$^{b,e}$}, Daniel Wolf\textit{$^{b}$}, Christian K. Müller\textit{$^{a}$}}$^{\ast}$ \\%

\includegraphics{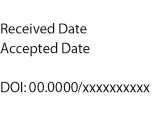} & \noindent\normalsize{This study examines how deposition potential influences the composition and nanoscale structure of Prussian Blue (PB)/Prussian White (PW)-family films, providing an experimental basis for assessing their proposed functions. Films were grown directly on Au/Cr/Si in a single potentiostatic step from a common aqueous bath at $-0.20$ to $+0.40$\,V versus a saturated calomel electrode, with a fixed deposition charge of 10\,mC. Electron microscopy, elemental analysis, X-ray photoelectron spectroscopy, and Raman spectroscopy revealed potential-dependent chemistry and nonmonotonic morphology. EDS indicated greater K incorporation in lower-potential deposits, whose Fe~2p and Raman features were consistent with more reduced, PW-like environments. Surface K/Fe ratios varied nonmonotonically from 1.129 to 1.322 across the series, corresponding to 52.95--56.94\% K within the detected K+Fe population. Larger faceted crystallites at $+0.30$\,V contrasted with finer pyramids and cracks at $+0.40$\,V. Selected PB-like and PW-like regions had local cubic lattice parameters of $(10.31\pm0.10)$\,\AA\ and $(10.36\pm0.10)$\,\AA, respectively; their overlapping uncertainties prevent discrimination by lattice spacing alone. Cross-sectional maps showed broadly uniform framework-element distributions within locally continuous coatings. Together, these observations relate deposition conditions to contrasting chemical and morphological starting states without establishing bulk stoichiometry or phase purity. A literature-grounded outlook identifies ion storage, sensing, electrochromism, resistive switching, and ion handling as research opportunities requiring functional validation in these films.} \\

\end{tabular}

 \end{@twocolumnfalse} \vspace{0.6cm}

  ]

\renewcommand*\rmdefault{bch}\normalfont\upshape
\rmfamily
\section*{}
\vspace{-1cm}

\footnotetext{\textit{$^{a}$~Faculty of Physical Engineering/Computer Sciences, University of Applied Sciences Zwickau, 08056 Zwickau, Germany. }}
\footnotetext{\textit{$^{b}$~Leibniz Institute for Solid State and Materials Research Dresden, 01069 Dresden, Germany. }}
\footnotetext{\textit{$^{c}$~Institute of Condensed Matter and Nanosciences (ICMN), Université catholique de Louvain (UCLouvain), Louvain-la-Neuve, B-1348, Belgium }}
\footnotetext{\textit{$^{d}$~Dresden Center for Nanoanalysis, Center for Advancing Electronics Dresden, TU Dresden, 01069 Dresden, Germany. }}
\footnotemark[5]\footnotetext{\textit{$^{e}$~Institute of Solid State and Materials Physics, TU Dresden, 01069 Dresden, Germany. }}
\footnotetext{\textit{$^{\ast}$~Corresponding author, E-mail: Christian.Mueller.1@whz.de}}

\section{Introduction}\label{sec:introduction}
Prussian Blue (PB), iron(III) hexacyanoferrate(II), is one of the earliest synthetic pigments and a reference material in cyanide-bridged coordination chemistry.\cite{Koncki2002_PrussianBlueSensors,Buser1977,Garcia2026_PBAPowders} Its mixed-valence framework contains C-coordinated Fe(II) and N-coordinated Fe(III), conventionally represented as Fe$^{\mathrm{II}}$--C$\equiv$N--Fe$^{\mathrm{III}}$. Variations in alkali-metal content, hexacyanoferrate vacancies, and water occupancy give rise to compositions commonly represented by the limiting formulas $\mathrm{Fe_4[Fe(CN)_6]_3\cdot xH_2O}$ and $\mathrm{KFe[Fe(CN)_6]}$. The historical terms ``insoluble'' and ``soluble'' PB describe compositional families rather than the actual aqueous stability of individual films. Reduction of N-coordinated Fe(III), accompanied by charge-compensating cation incorporation, produces more reduced Prussian White (PW)-like environments, with $\mathrm{K_2Fe[Fe(CN)_6]}$ representing an idealized limiting composition.\cite{Faita2022_PrussianWhite_ResistiveSwitching}

\begin{figure*}
 \centering
 \includegraphics[height=5cm]{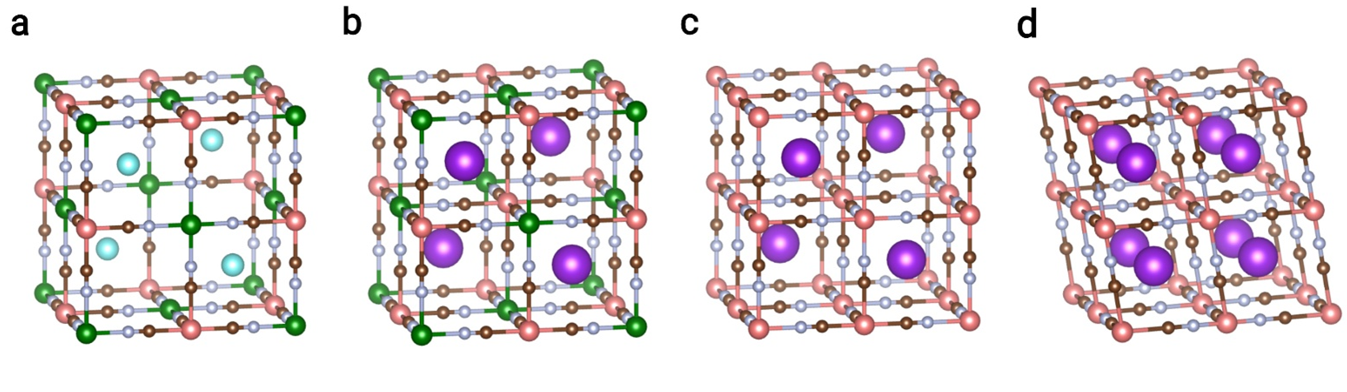}
 \includegraphics[height=1.7cm]{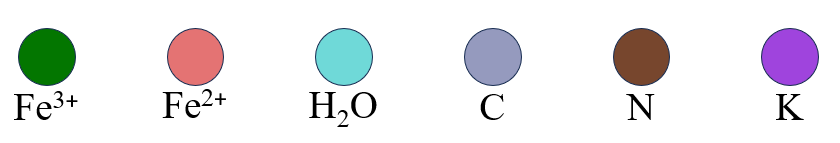}
 \caption{Schematic structural models: (a) vacancy-containing, alkali-poor PB; (b) alkali-containing PB; (c) cubic PW; and (d) a rhombohedrally distorted PW model. Coordinated water molecules (H$_2$O) are drawn only in (a) for clarity. These illustrations provide structural context and do not establish the space group, water content, or phase fractions of the present films.}
 \label{fig:pb_crystal_structure}
\end{figure*}

PB and PW belong to the broader family of Prussian Blue analogues (PBAs), generally represented as $A_xM[M'(\mathrm{CN})_6]_{1-y}\square_y\cdot n\mathrm{H_2O}$, where $A$ denotes an interstitial cation, $M$ and $M'$ are transition metals, and $\square$ represents a hexacyanometallate vacancy. Their open-framework structures accommodate ionic insertion and extraction, while vacancies, metal substitution, and synthesis conditions influence structural stability, charge transport, and electrochemical properties.\cite{Simonov2020,Garcia2025,Quispe2021,Serrano2021} Recent solution-based studies of Fe-, Co-, and Ni-based PBAs further demonstrate the influence of composition and preparation conditions on their structural, spectroscopic, and thermal characteristics.\cite{Garcia2026_PBAPowders} Although cubic $Fm\overline{3}m$ structures are frequently reported, lower-symmetry arrangements may emerge depending on composition and oxidation state. In particular, rhombohedral/cubic transformations observed in sodium-containing PW batteries cannot be directly extrapolated to potassium-containing films.\cite{Nielsen2024_PW_FullCells_OperandoXRD} Figure~\ref{fig:pb_crystal_structure} illustrates possible structural arrangements rather than experimentally determined phase fractions. Recent advances in defect-engineered piezoelectric nanomaterials and fluorescent carbon dots illustrate how composition, structural defects, and electronic properties can be tailored for emerging sensing, imaging, and biomedical applications.\cite{Ullah2025Small,Hasan2026JNB,Madni2026CBI,Roy2025FrontChem}

The combination of an open framework, reversible Fe-based redox chemistry, and coupled ionic-electronic transport makes PB and PW attractive for several technological applications. Their ability to accommodate alkali-metal ions, particularly Na$^+$ and K$^+$, provides opportunities for rechargeable battery electrodes, pseudocapacitive energy storage, and hybrid supercapacitors. Depending on the electrochemical configuration and operating potential, these materials may participate in anodic or cathodic processes. Furthermore, reversible changes in oxidation state and optical absorption enable electrochromic applications, including smart windows and displays. Their electrochemical activity is also relevant to chemical sensing, while ion-mediated electronic transport offers opportunities for resistive-switching memories.\cite{Koncki2002_PrussianBlueSensors,Faita2022_PrussianWhite_ResistiveSwitching} This notation describes chemical degrees of freedom; it is not a composition refinement of the present samples. Vacancy connectivity can influence accessible transport pathways, while metal substitution changes bonding and electrochemical response.\cite{Simonov2020,Garcia2025,Quispe2021} Cubic $Fm\overline{3}m$ models and lower-symmetry structures are both reported, with symmetry depending on composition and state of charge.

Electrodeposition offers a simple, cost-effective, and versatile solution-based approach for fabricating PB/PW thin films directly onto conductive substrates. By adjusting the deposition potential, it is possible to influence potassium incorporation, local chemical environments, and film morphology, providing a practical route for tailoring material properties. In this work, we systematically investigate the relationship between deposition potential, chemical composition, morphology, and local crystallographic structure through complementary structural, spectroscopic, and microscopic characterization. Particular attention is given to the evolution between PB-like and PW-like chemical environments, without assuming the formation of phase-pure compounds. Our findings establish a materials-level foundation for the future development of PB/PW-based electrodes for energy storage, electrochromic, sensing, and electronic applications.

These coupled chemical and structural variables motivate applications in alkali-ion storage, peroxide sensing, electrochromism, and resistive switching. Published devices establish the relevance of PB-family materials in each area, but the requirements differ: an insertion electrode needs reversible ion inventory, a sensor needs selective interfacial chemistry, and a memory element needs reproducible electrical states and contact control.\cite{Lemaire2022_Interface,Karyakin1999_ArtificialPeroxidase,Assis2016_PB_ElectrochromicDevices,Avila2020_ResistiveSwitching_PrussianBlue} Relating a preparation condition to one of these functions therefore requires more than a structural assignment.

PBAs can be prepared by precipitation and assembled into films by several routes.\cite{Kjeldgaard2021_StrategiesPBA_Synthesis,Zhang2024_YtubePBAs_SIB,Camacho2021_NaRichPBA_SynthesisImpact,Li2025_LowDefectPrussianBlue} Electrodeposition is useful for studying interfaces because it grows the coordination material directly on a conducting support without a powder-slurry binder. Potential, solution chemistry, and growth charge provide experimentally accessible preparation variables. Earlier work has already established electrodeposited PW films and composition-dependent morphologies in metal-substituted HCFs.\cite{Faita2022_PrussianWhite_ResistiveSwitching,Garcia2025}

Here, the contribution is a complementary surface and cross-sectional characterization of an iron-only potential series, including surface ratios and Fe chemical environments, cyanide vibrations, local crystallography, and nanoscale elemental distributions. Films were prepared at $-0.20$ to $+0.40$\,V versus SCE at a common growth charge. ``Single-step'' denotes formation of the PB/PW coating in one potentiostatic deposition from the mixed bath; it does not include fabrication of the Au/Cr/Si support or imply that a complete device requires only one processing step. Section~\ref{sec:results} establishes the observations and their limits. Section~\ref{sec:applications-outlook} then uses those specific results to assess literature-supported application hypotheses and define the measurements needed to test them.
\section{Experimental}\label{sec:experimental}
\subsection{Materials and electrodeposition}\label{sec:deposition}
PB/PW-family films were deposited in a three-electrode configuration at room temperature using an Ivium CompactStat potentiostat (Ivium Technologies, Eindhoven, The Netherlands). The working electrodes were Au/Cr/Si substrates of 1\,cm $\times$ 1.5\,cm, the counter electrode was platinum foil, and the reference was a saturated calomel electrode (SCE). All deposition potentials in this article are referenced to SCE. The reference electrode was calibrated by its manufacturer before use; an additional in-cell calibration is not documented in the preparation notes. The supports consisted of 50\,nm Au thermally evaporated onto a 5\,nm Cr adhesion layer on (100)-oriented Si, at a base pressure of $10^{-5}$\,Pa. An adhesive tape mask defined a circular exposed area of 0.5\,$\mathrm{cm^2}$.

Substrates were cleaned sequentially in ultrasonic baths of acetone and distilled water, thoroughly rinsed with distilled water, and dried with nitrogen before deposition. The final aqueous electrolyte contained 0.25\,mM K$_3$Fe(CN)$_6$ (ACS, $>99\%$), 0.25\,mM FeCl$_3$ (ACS, 98\%), 1.0\,M KCl (ACS, 99--100.5\%), and 5.0\,mM HCl (ACS, 37\% reagent), all from Sigma Aldrich, Darmstadt, Germany. The reported bath pH was approximately 2.2. The supporting KCl/HCl solution was prepared under atmospheric conditions and divided into two equal-volume portions for separate dissolution of FeCl$_3$ and K$_3$Fe(CN)$_6$; these portions were mixed immediately before deposition to limit premature solution reaction. The preparation notes specify 1\,mol KCl and 5\,mmol HCl at this stage, but the associated batch volume requires confirmation against the final concentrations stated above. Cleaning times and stock-solution volumes are not specified in the available record.

Potentiostatic deposition at $-0.20$ to $+0.40$\,V was stopped at a total charge of 10\,mC per sample, corresponding to a nominal growth charge of 20\,mC\,cm$^{-2}$. However, equal growth charge does not establish equal deposited mass, thickness, or Faradaic efficiency. After deposition the films were removed immediately, rinsed thoroughly with distilled water, dried with nitrogen, and stored in air between measurements. For each potential at least three samples were produced to prove the reproducibility of the deposition experiments.

\subsection{Film characterization}\label{sec:characterization}
Morphology and elemental composition were examined by field-emission SEM (TESCAN CLARA, Brno, Czech Republic) with an Ultim Max 65 SDD EDS detector (Oxford Instruments, Wiesbaden, Germany), at an accelerating voltage of 10\,kV. EDS spectra were quantified using the same standardless routine in Oxford Instruments AZtec for all samples. Raman spectra were acquired using the SEM-integrated WITec RISE system (Ulm, Germany), a 532\,nm laser, and a power of 0.9\,mW.

Surface composition was measured by XPS using a Genesis photoelectron spectrometer (Physical Electronics, USA) with monochromatized, micro-focused Al~K$\alpha$ radiation at 50\,W. Conductive clips secured the samples, and no charge stabilization was applied under these conditions. The analysis-chamber pressure was approximately $10^{-6}$\,Pa, and the angle between the surface normal and analyzer-lens axis was $0^{\circ}$. Spectra were analyzed in CasaXPS version 2.2.24. The K/Fe ratios and K--Fe-normalized fractions in Table~\ref{tab:xps-ratios-total} are reported as means and sample standard deviations from four analyses per potential. This count describes analyses, rather than an independently established number of deposited films. The interpretation and provenance of the other quantification entries are discussed in Section~\ref{sec:xps}.

Cross-sectional lamellae of approximately $15\,\mu$m $\times$ $10\,\mu$m $\times$ $50\,\mathrm{nm}$ were prepared by FIB milling. Bright-field TEM (BF-TEM) and HRTEM used a double-corrected FEI Titan$^3$ 80--300 microscope (Thermo Fisher Scientific, USA) at 300\,keV; the BF-TEM objective aperture had a diameter of $50\,\mu$m. A carbon sputter layer followed by platinum protected the films during lamella preparation.

STEM--EELS of the PB-like film was performed on a Hitachi HF3300\,S microscope at 300\,keV equipped with a CEFID post-column imaging energy filter\cite{Kahl2019} and TVIPS XF416 detector. The selected energy dispersion covered 1024\,eV with an energy resolution of approximately 1.3\,eV. PW was examined using a JEOL JEM F-200C microscope at 200\,keV with a Gatan GIF Continuum ER filter. Elemental maps of Fe, K, C, N, and O were obtained with background fitting and core-loss quantification in Gatan DigitalMicroGraph; the C/K separation limitation is retained in interpreting these maps. 

HAADF-STEM imaging and STEM--EDS of both films were performed on a Talos F200i (Thermo Fisher Scientific, USA) at 200\,kV with a Dual-X EDS detector system, in nanoprobe mode with a 70\,$\mu$m C2 aperture and a convergence semi-angle of 10.5\,mrad. Spectrum images were acquired with a dwell time of 12.5\,$\mu$s per pixel and summed over multiple frames: for the PB-like film at a screen current of 407\,pA and a pixel size of 5.72\,nm (165 frames, about 10~min), and for the PW-like film at 216\,pA and 0.71\,nm (227 frames, about 24~min). The resulting cumulative electron doses are about $1.6\times10^{3}$~e$^{-}$\,\AA$^{-2}$ (PB-like) and $7.6\times10^{4}$~e$^{-}$\,\AA$^{-2}$ (PW-like). Spectrum images were quantified in Velox by the standardless Cliff--Lorimer method with theoretical $k$-factors (Brown--Powell ionization cross-sections), a second-order multi-polynomial background fitted in automatically selected windows, with C, N, O, K and Fe and normalized to 100~at.\%; K and Fe were quantified from their K$\alpha$ lines. Line profiles from the Si substrate to the top of the lamella (Fig.~\ref{fgr:EDS_PB}c and \ref{fgr:EDS_PW}c) were used to show the layer sequence. For quantification, profiles were drawn parallel to the Au interface within the film, integrating perpendicular to the line over a window of 208 pixels (1.19\,$\mu$m) for the PB-like and 133 pixels (95\,nm) for the PW-like film, placed within the film interior so that the Au and protection-layer interfaces were excluded; K/(K+Fe) was evaluated point by point along these profiles. The stated uncertainty is the standard deviation of the point-by-point K/(K+Fe) values along each profile; it reflects counting noise and local compositional variation within one lamella, not the reproducibility between films. Light-element values are treated qualitatively.

\section{Results and discussion}\label{sec:results}
\subsection{Morphology and elemental composition}\label{sec:morphology}
\begin{figure*}[!b]
 \centering
 \includegraphics[height=9.2cm]{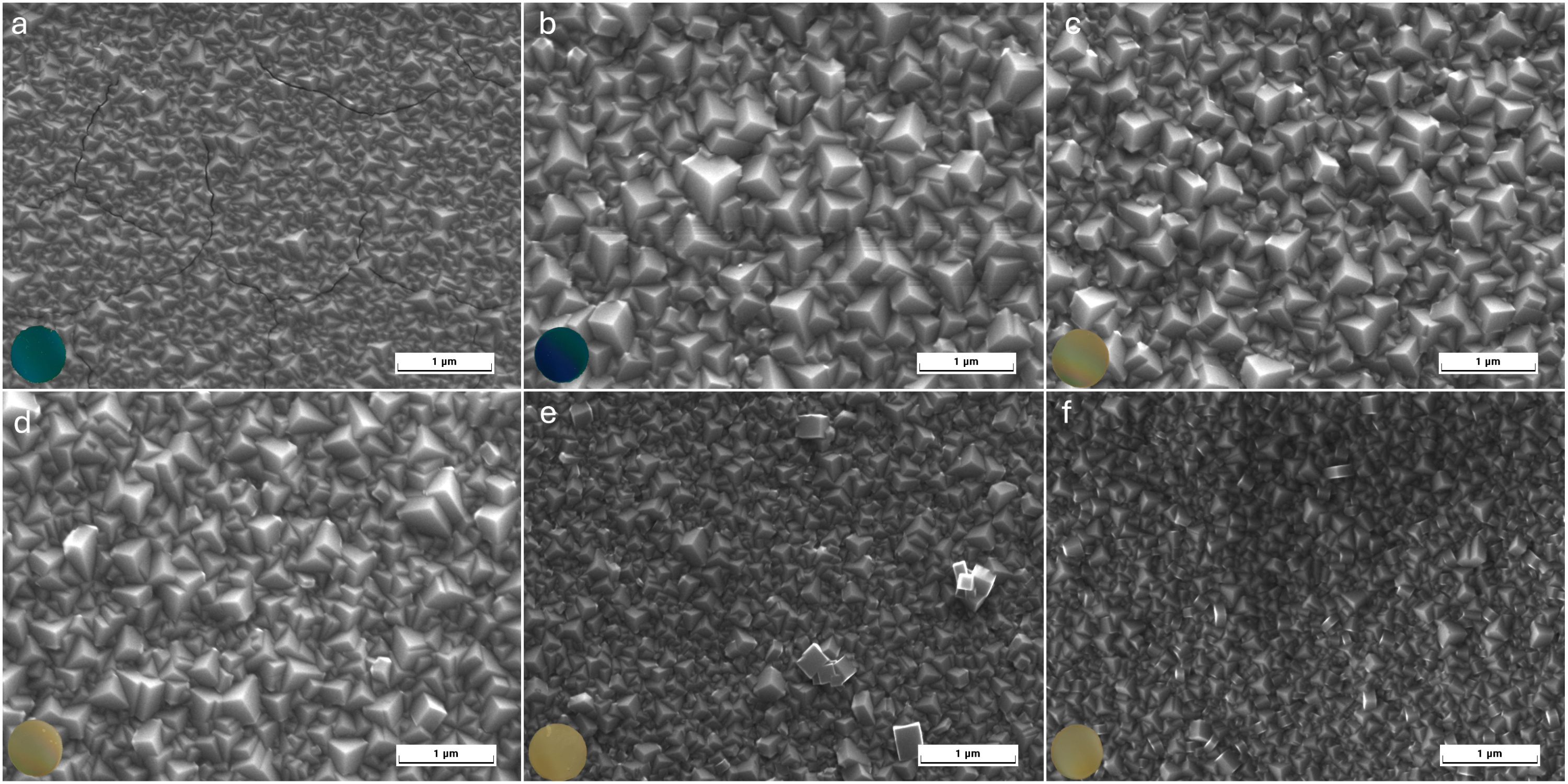}
 \caption{SEM images of films deposited at (a) 0.4, (b) 0.3, (c) 0.2, (d) 0.1, (e) $-0.1$, and (f) $-0.2$~V versus SCE. Insets show photographs of the as-deposited films. The photographs compare different specimens; they are not an electrochromic switching experiment.}
 \label{fgr:SEM of PB_PW overview}
\end{figure*}

SEM reveals a nonmonotonic morphological response to deposition potential (Figure~\ref{fgr:SEM of PB_PW overview}). The films have a typical thickness of approximately 0.2--1.0\,$\mu$m. The 0.40\,V PB-like film contains densely packed pyramidal crystallites, with lateral dimensions reported below 200\,nm, and visible cracks.  At 0.30\,V, larger, predominantly cubic faceted crystallites are observed without visible cracks in the examined field. At 0.10\,V, the PW-like film contains pyramidal grains with reported dimensions of approximately 0.1--0.6\,$\mu$m. These dimensions are also in agreement with x-ray diffraction experiments (not shown).

Changes in nucleation and growth rates offer a possible explanation for the contrasting morphologies, but deposition-current transients and grain statistics are needed to separate those processes. Faceted shapes are compatible with the cyanoferrate framework; earlier characterization of PW-film provides a related morphological comparison \cite{Faita2022_PrussianWhite_ResistiveSwitching}. Morphology alone does not establish an exposed-plane distribution or a quantitatively preferred (111) texture; local crystallographic observations are discussed in Section~\ref{sec:tem}. The cracks at 0.40\,V could reflect growth or drying stresses, but stress and mechanical adhesion were not measured. Small unaligned clusters at $-0.10$ and $-0.20$\,V could represent different nucleation/coalescence histories or local heterogeneity; their presence does not by itself establish a second phase.

\begin{figure*}[t!]
 \centering
 \includegraphics[width=\linewidth]{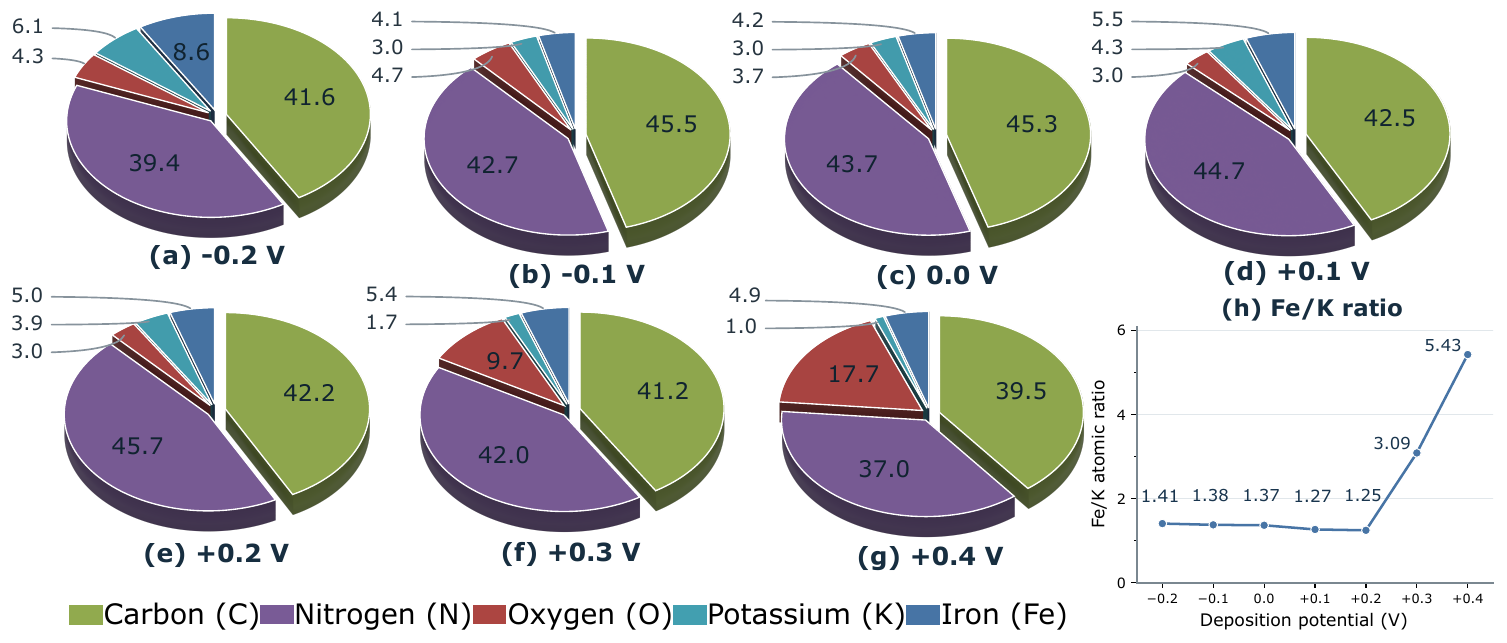}
 \caption{EDS elemental compositions (at.\%) of films deposited at (a) $-0.20$, (b) $-0.10$, (c) 0.00, (d) $+0.10$, (e) $+0.20$, (f) $+0.30$, and (g) $+0.40$\,V versus SCE. Colours identify C, N, O, K, and Fe as indicated in the legend. (h) Fe/K atomic ratio as a function of deposition potential.}
 \label{fgr:EDS_results}
\end{figure*}

EDS indicates lower K incorporation in the 0.40 and 0.30\,V PB-like films (1.0 and 1.7~at.\%, respectively) than in lower-potential films (up to 6.1~at.\%; Figure~\ref{fgr:EDS_results}a--g). Fe varies more moderately. The EDS Fe/K ratio decreases from approximately 5.4 at 0.40\,V to approximately 1.4 at $-0.20$\,V (Figure~\ref{fgr:EDS_results}h), consistent with greater potassium incorporation at lower potential. The oxygen signal is generally lower in the lower-potential series than in the high-potential PB-like films, but does not provide a direct water-content measurement: adsorbed species, contamination, and sampling of adjacent material can also contribute. Hydration and hexacyanoferrate-vacancy concentrations therefore remain undetermined.

\subsection{XPS spectroscopy}\label{sec:xps}
Table~\ref{tab:xps-ratios-total} summarizes surface atomic ratios, K--Fe-normalized percentages, and elemental sums. K/Fe is approximately 1.26--1.33 between $-0.20$ and $+0.20$\,V, then falls to 0.593 at $+0.30$\,V and 0.261 at $+0.40$\,V. The largest sampled compositional change is between $+0.20$ and $+0.30$\,V; the lower-potential series is not strictly monotonic. XPS K/Fe and EDS Fe/K have reciprocal definitions and different sampling depths, so only their qualitative potential trends should be compared directly. Surface ratios are not sufficient to assign exact bulk formulas or vacancy fractions.

Because the denominator includes both framework Fe sites, ideal $\mathrm{K_2Fe^{II}[Fe^{II}(CN)_6]}$ has K/Fe = 1. Surface ratios above this value cannot be interpreted directly as that bulk formula; surface enrichment, residual potassium-containing species, or quantification bias are possible contributions. The greater measured K content is therefore a compositional observation rather than proof of a fully occupied, vacancy-free PW lattice.

The table's numerical entries are retained as reported. Only 13 individual analyses are present in the supplied quantification export, despite the reported count of four analyses at each potential. In addition, some edited SDs for the complementary K--Fe percentages differ although their sum is exactly 100\% by definition; those SDs should be identical for the same analysis set. The full individual results and the provenance of these uncertainty entries require verification. Element-sum SDs from 0.00 to $+0.40$\,V remain explicitly labelled pooled estimates. These limitations preclude using the tabulated uncertainties to claim statistically significant differences or reproducibility across the entire series.

\begin{table*}[!t]
\centering
\caption{XPS-derived atomic ratios, K--Fe-normalized composition,
and elemental totals as a function of deposition potential.
Updated K/Fe ratios and K--Fe fractions are reported as
mean $\pm$ sample standard deviation from four analyses
per potential ($n=4$).}
\label{tab:xps-ratios-total}

\small
\renewcommand{\arraystretch}{1.15}
\setlength{\tabcolsep}{4pt}

\resizebox{\linewidth}{!}{%
\begin{tabular}{@{}rcrrrrrrr@{}}
\toprule
\shortstack{$E_{\mathrm{dep}}$\\(V)}
& $n$
& K/Fe
& N/Fe
& C/Fe
& O/Fe
& \shortstack{$100\,\mathrm{K}/(\mathrm{K}+\mathrm{Fe})$\\(\%)}
& \shortstack{$100\,\mathrm{Fe}/(\mathrm{K}+\mathrm{Fe})$\\(\%)}
& \shortstack{Element sum\\(at.\%)} \\
\midrule

$-0.20$ & 4
& $1.262 \pm 0.040$
& $3.640 \pm 0.041$
& $6.283 \pm 0.211$
& $0.919 \pm 0.092$
& $55.79 \pm 0.78$
& $44.21 \pm 0.78$
& $99.98 \pm 0.15$ \\

$-0.10$ & 4
& $1.322 \pm 0.033$
& $3.685 \pm 0.034$
& $6.671 \pm 0.152$
& $1.222 \pm 0.038$
& $56.94 \pm 0.60$
& $43.06 \pm 0.60$
& $100.03 \pm 0.06$ \\

$0.00$ & 4
& $1.271 \pm 0.063$
& $3.714 \pm 0.038$
& $6.729 \pm 0.184$
& $1.143 \pm 0.070$
& $55.93 \pm 1.22$
& $44.07 \pm 1.22$
& $99.83 \pm 0.11$ \\

$+0.10$ & 4
& $1.129 \pm 0.102$
& $3.681 \pm 0.030$
& $6.236 \pm 0.180$
& $1.083 \pm 0.075$
& $52.95 \pm 2.33$
& $47.05 \pm 2.33$
& $100.00 \pm 0.11$ \\

$+0.20$ & 4
& $1.211 \pm 0.133$
& $3.573 \pm 0.031$
& $6.080 \pm 0.181$
& $1.013 \pm 0.075$
& $54.65 \pm 2.71$
& $45.35 \pm 2.71$
& $100.00 \pm 0.11$ \\

$+0.30$ & 4
& $1.248 \pm 0.088$
& $4.695 \pm 0.040$
& $8.390 \pm 0.174$
& $1.847 \pm 0.078$
& $55.47 \pm 1.73$
& $44.53 \pm 1.73$
& $100.10 \pm 0.11$ \\

$+0.40$ & 4
& $1.184 \pm 0.050$
& $5.000 \pm 0.036$
& $12.935 \pm 0.188$
& $2.565 \pm 0.080$
& $54.19 \pm 1.05$
& $45.81 \pm 1.05$
& $100.10 \pm 0.11$ \\

\bottomrule
\end{tabular}%
}

\par\vspace{4pt}
\begin{minipage}{\linewidth}
\footnotesize
K/Fe and both normalized fractions were calculated separately
for each analysis before determining their means and sample
standard deviations. 

\end{minipage}
\end{table*}

\begin{figure*}[!t]
 \centering
 \includegraphics[width=0.80\textwidth, keepaspectratio,trim={0 0 0 50bp},clip]{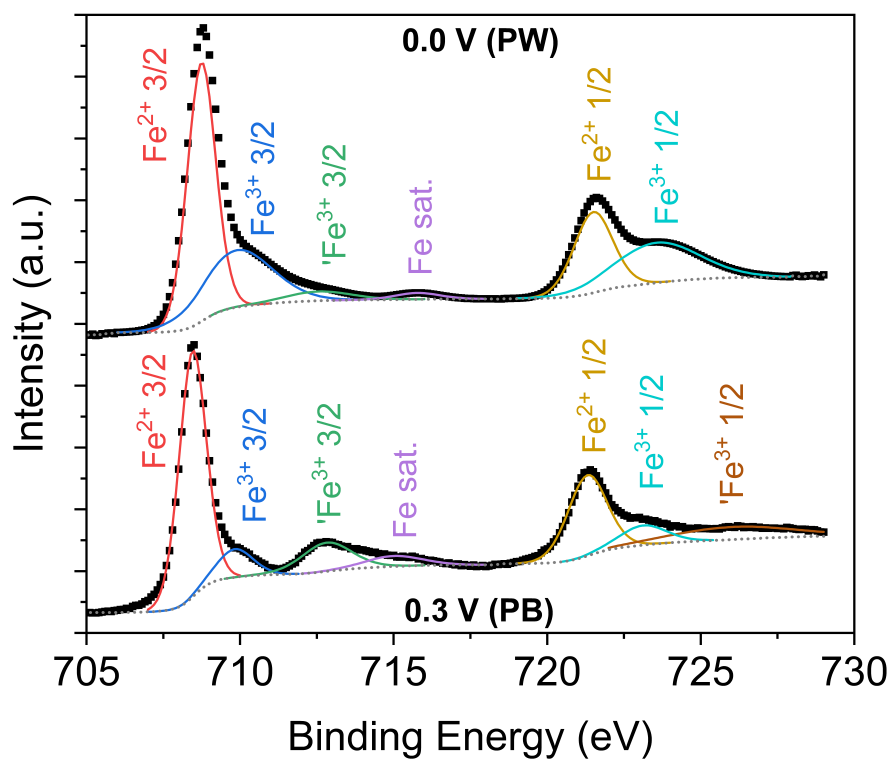}
 \caption{Fe~2p XPS spectra and the reported component fits over 705--730\,eV for the 0.0\,V PW-like and 0.3\,V PB-like films. Component labels describe the adopted fit and do not provide independently validated oxidation-state fractions.}
 \label{fgr:XPS results}
\end{figure*}

The survey spectra contain C near 285\,eV, K near 294 and 378\,eV, N near 398\,eV, O near 532\,eV, and Fe near 709 and 722\,eV (Figure~S1, Supporting Information), consistent with the elemental constituents of cyanoferrate films. Similar surface analyses are reported for related systems.\cite{Su2017PrussianBlueGrapheneCF,Yan2017_PrussianBlueCathode,Garcia2025} The weaker K signal above 0.20\,V supports the compositional trend, but does not separately identify soluble/insoluble PB fractions. A signal near 689\,eV is attributed to F; chamber-derived contamination is a possible origin, not an established cause. Consequently, the survey does not demonstrate chemically clean surfaces.

The Fe~2p spectra of the 0.00\,V PW-like and 0.30\,V PB-like films contain the expected spin--orbit regions near 708--710 and 720--724\,eV (Figure~\ref{fgr:XPS results}). Literature comparison places Fe(II) $2p_{3/2}$ near 708.4\,eV and ferric cyanoferrate contributions around 709.8--710.5\,eV.\cite{Zhang2022_LithiatedPrussianBlueAnalogues} The reported fits assign components near 708.5--708.7\,eV to Fe(II) and 709.8--709.9\,eV to Fe(III), with the $2p_{1/2}$ region approximately 13\,eV higher. These chemical assignments are consistent with more reduced and mixed-valence environments, respectively, but peak positions and widths do not establish a Fe(II)/Fe(III) area ratio or spin-state population.

For the 0.00\,V film, the main component at 708.7\,eV has a reported FWHM of approximately 1.1\,eV; a broader component at 709.9\,eV has a FWHM of approximately 2.7\,eV. The fit also contains features at 712.7\,eV and approximately 715.7\,eV, labelled ferric and ferrous satellite contributions, respectively. In the $2p_{1/2}$ region, the reported components are 721.5\,eV (FWHM approximately 1.4\,eV) and 723.4\,eV (approximately 3.5\,eV). Surface oxidation, defects, and overlapping multiplets are possible contributions to broad features. The ferric contribution cannot be dismissed solely because a reduced PW assignment is expected; the integrated areas, background, constraints, and residuals are needed to assess that interpretation. General Fe~2p multiplet analyses provide fitting context,\cite{Grosvenor2004_InvestigationMultipletSplitting}but do not independently validate the cyanoferrate-specific decomposition.

For the 0.30\,V film, the reported Fe(II) and Fe(III) $2p_{3/2}$ components lie at 708.5\,eV (FWHM approximately 1.0\,eV) and 709.8\,eV (approximately 1.4\,eV). The latter falls within the 709.8--710.0\,eV range discussed for ferric cyanoferrates.\cite{Zhang2022_LithiatedPrussianBlueAnalogues} Its narrower width than the fitted ferric contribution in the PW-like film could reflect a less distributed chemical environment, but is not by itself a measure of phase purity. Corresponding $2p_{1/2}$ components are reported at 721.3 and 723.1\,eV; an additional broad feature at 726\,eV (FWHM approximately 4.7\,eV) was included as a ferric multiplet-related contribution.

The PB-like fit contains a satellite near 712.8\,eV. Ferric-related satellites have been reported at 712.2 and 713.9\,eV in PB/PANI-containing coatings and near 712.6\,eV in PB nanoparticles.\cite{Wang2022_PhyticAcidDopedPolyaniline,Yang2021_GreenRoutePrussianBlueNPs} For comparison, a satellite near 718.8\,eV, roughly 7--8\,eV above a main feature near 711\,eV, is discussed for Fe(III) oxides.\cite{Yamashita2008_AnalysisXPSFeOxides} Ligand covalency, multiplet structure, and charge-transfer final states affect these comparisons. Because the PW-like fit itself includes the 712.7\,eV feature, the present data do not support an assertion that the nearby satellite is strictly absent in PW. This resolves the contradictory presence/absence statements without changing any reported fit energies. Quantitative oxidation-state fractions and high-spin populations require a validated fit and complementary spectroscopy.

\subsection{Raman spectroscopy}\label{sec:raman}
\begin{figure}[h!]
 \centering
 \includegraphics[width=1\linewidth]{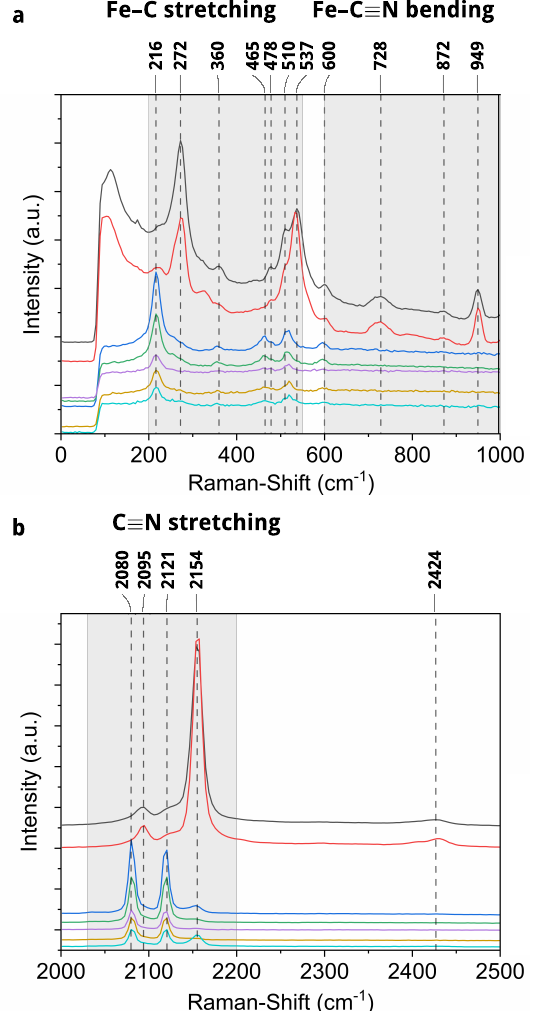}
 \caption{Raman spectra of electrodeposited Prussian Blue (PB) and Prussian White (PW) thin films prepared at different deposition potentials. (a) Low-wavenumber region showing characteristic Fe--C stretching and Fe--C$\equiv$N bending vibrations. (b) High-wavenumber region showing C$\equiv$N stretching vibrations. No Raman-active bands were reported between 1000 and 2000~cm$^{-1}$. Colours represent deposition potentials versus SCE: grey, $+0.40$ V; red, $+0.30$ V; blue, $+0.20$ V; green, $+0.10$ V; purple, $0.00$ V; ochre, $-0.10$ V; and cyan, $-0.20$ V.}
 \label{fig:raman_pb_pw}
\end{figure}

The spectra in Figure~\ref{fig:raman_pb_pw} retain cyanide stretching and lower-wavenumber framework modes across the potential series. Literature PB bands at 2094 and 2154\,cm$^{-1}$, and Fe--C-related bands at 278 and 535\,cm$^{-1}$, provide comparison points.\cite{Samain2013,Barsan2011} The PB-like films show a strong 2154\,cm$^{-1}$ feature and a weaker band near 2095\,cm$^{-1}$. A shoulder near 2121\,cm$^{-1}$ could include a different cyanoferrate environment or precursor-related contribution; it is not a unique fingerprint of an independently identified PW phase.\cite{Samain2014} The reported feature at 2424\,cm$^{-1}$ remains unassigned: a combination/overtone or measurement-related contribution is possible, but a fundamental PB vibration is not established.

In the approximately 400--1000\,cm$^{-1}$ region, bands near 476\,cm$^{-1}$ (labelled 478\,cm$^{-1}$ in the supplied figure), 728, 872, and 949\,cm$^{-1}$ were attributed provisionally to framework bending, deformation, or combination modes. Comparison bands near 483 and 712\,cm$^{-1}$ have also been discussed in Raman studies involving PB pigments.\cite{Steger2019} Specific assignments, especially 872 and 949\,cm$^{-1}$, should remain tentative. Peaks at 272 and 537\,cm$^{-1}$ are compatible with the literature Fe--C bands at 278 and 535\,cm$^{-1}$; the feature at 360\,cm$^{-1}$ can be compared with a reported deformation near 338\,cm$^{-1}$. Bands near 510 and 476\,cm$^{-1}$ and the weak 216\,cm$^{-1}$ signal provide additional lattice/Fe--ligand information.\cite{Barsan2011,Steger2019}

At lower deposition potentials, the 2154\,cm$^{-1}$ feature becomes weaker relative to the bands at 2121 and 2080\,cm$^{-1}$. This change is consistent with a more reduced cyanoferrate environment, complementing the K-rich surface composition and the Fe~2p comparison. The correct PB connectivity is Fe$^{\mathrm{II}}$--C$\equiv$N--Fe$^{\mathrm{III}}$, whereas ideal PW contains predominantly Fe$^{\mathrm{II}}$--C$\equiv$N--Fe$^{\mathrm{II}}$ bridges. Literature symmetry labels such as A$_{1g}$ and E$_g$ do not by themselves resolve all overlapping environments in these films. The frequency downshift is compatible with changed Fe--CN bonding on reduction; differences in resonance enhancement and optical absorption also affect Raman intensities, so band intensity alone is not a quantitative valence fraction.

The PW-like spectra retain features around 510 and 465\,cm$^{-1}$, with weaker or unresolved 272 and 360\,cm$^{-1}$ modes and a weak signal near 216\,cm$^{-1}$. Broadening and small shifts could reflect a distribution of local environments. Loss of a visible band does not independently prove increased crystallographic symmetry, and persistence of CN bands does not demonstrate electronic conductivity or reversible electrochemical cycling. Together, Raman and XPS support a potential-dependent chemical evolution; they do not establish phase purity or device performance.

\subsection{Transmission electron microscopy}\label{sec:tem}

\begin{figure}[b!]
 \centering
 \includegraphics[width=\linewidth]{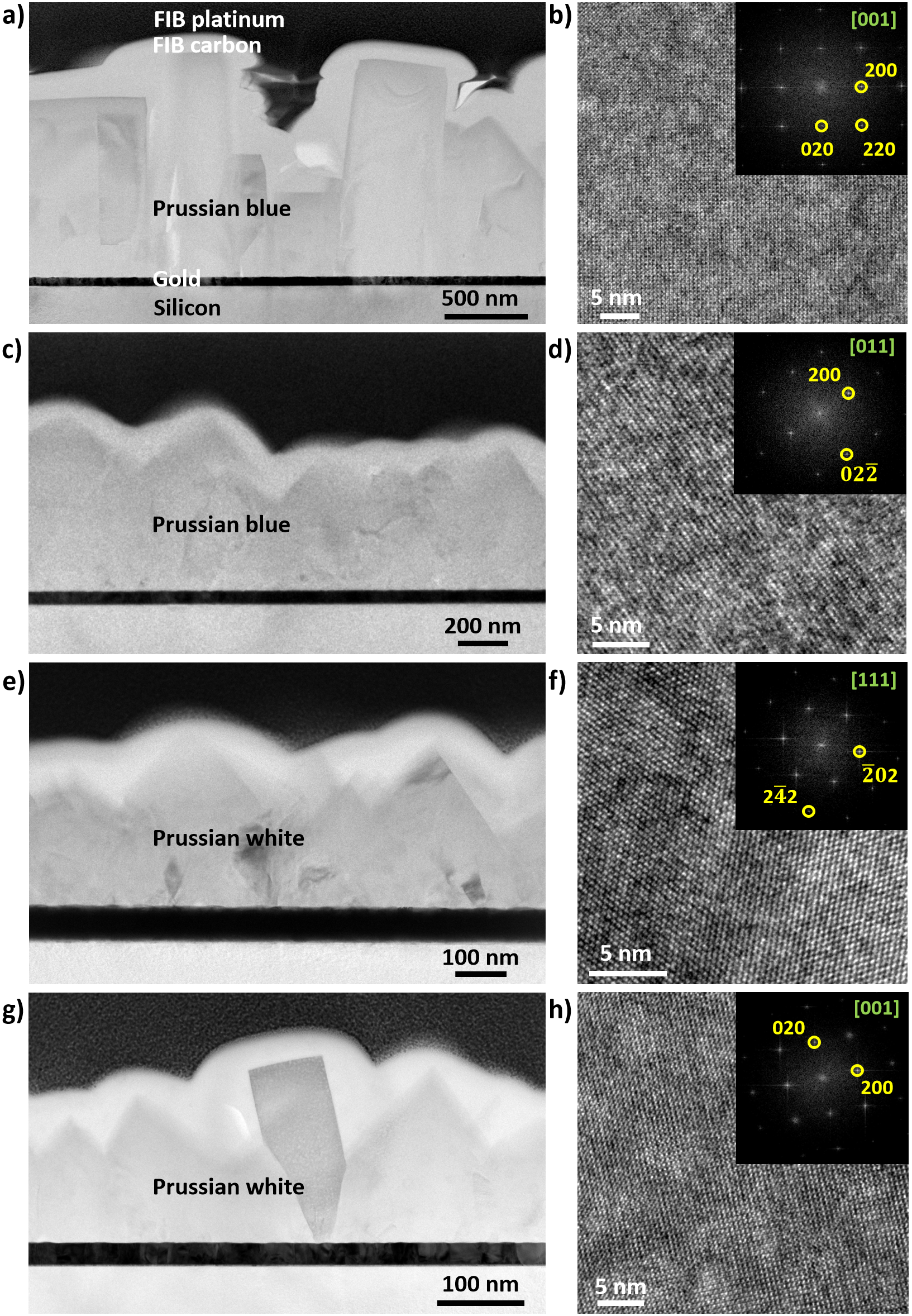}
 \caption{Cross-sectional TEM of the PB-like (0.30~V; a--d) and PW-like ($-0.10$~V; e--h) films. (a,c,e,g) BF-TEM overviews of FIB-prepared lamellae; dark-contrast regions are crystallites aligned with low-index zone axes. (b,d,f,h) Corresponding HRTEM images with Fourier transforms (insets), indexed to the [001], [011], [111] and [001] zone axes, respectively, consistent with a cubic framework in the examined regions.}

 \label{fgr:HRTEM}
\end{figure}

\begin{figure}[h!]
 \centering
 \begin{subfigure}{\linewidth}
    \centering
    \includegraphics[width=\linewidth, keepaspectratio]{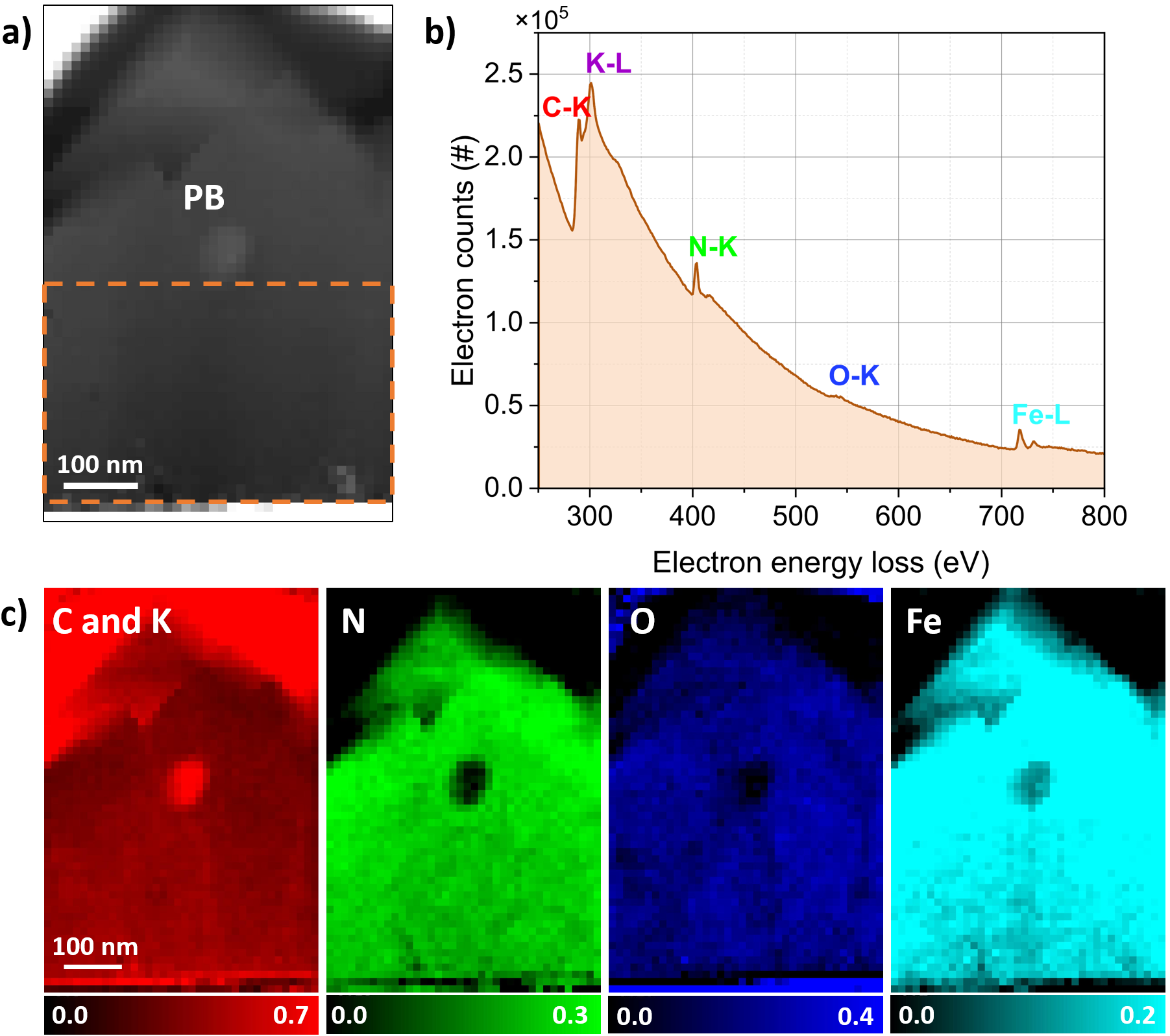}
  \end{subfigure}
 
  \begin{subfigure}{\linewidth}
    \centering
    \includegraphics[width=\linewidth, keepaspectratio]{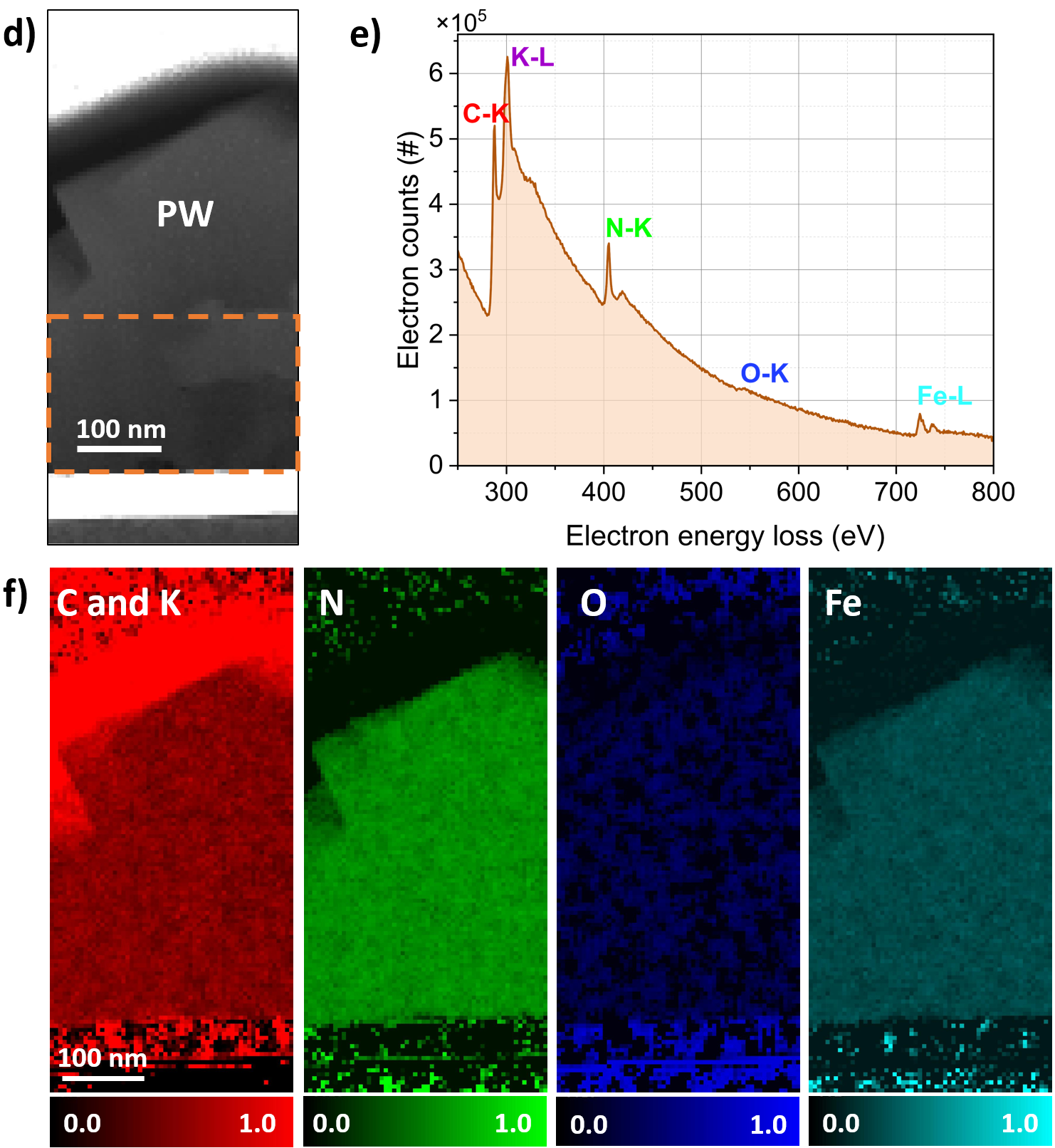}
  \end{subfigure}
  
 \caption{STEM--EELS of the PB-like (a--c, Hitachi HF3300S, 300~kV) and PW-like (d--f, JEOL JEM-F200, 200~kV) films. (a,d) HAADF-STEM images; orange boxes mark the regions from which the EEL spectra were acquired. (b,e) EEL spectra showing the C-K, K-L$_{2,3}$, N-K, O-K and Fe-L$_{2,3}$ edges. (c,f) Core-loss elemental maps; C and K are shown as a combined signal because their edges overlap. Because the data come from different instruments, map intensities are not compared between the films.}
 \label{fgr:EELS}
\end{figure}

\begin{figure*}[!h]
 \centering
    \includegraphics[width=0.85\linewidth, keepaspectratio]{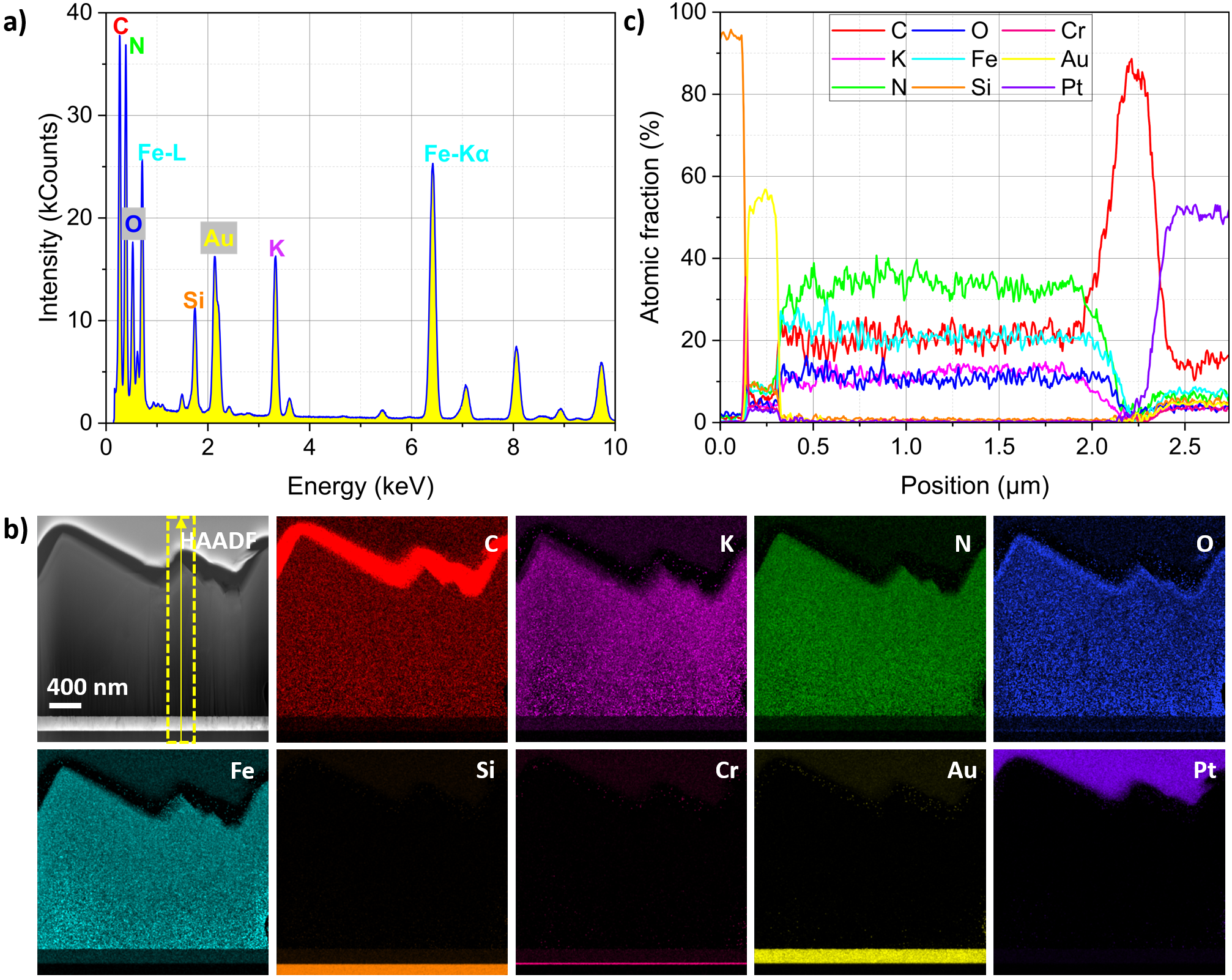}
   \caption{STEM--EDS of the PB-like film (0.30~V). (a) Spectrum from the analyzed region, showing C, K, N, O, Fe, Si, Cr, Au, and Pt. (b) HAADF-STEM image and elemental maps of the same elements across the lamella. (c) Line profile from the Si substrate to the top of the lamella along the yellow-marked region in (b).}
 \label{fgr:EDS_PB}
\end{figure*}

Cross-sectional BF-TEM, HRTEM, STEM--EELS, and STEM--EDS were performed on representative films deposited at 0.30\,V (PB-like) and $-0.10$\,V (PW-like). The aim was to determine which film properties accessible by TEM differ between the two films, given the gradual compositional evolution found by SEM--EDS, XPS and Raman spectroscopy (Sections \ref{sec:morphology}, \ref{sec:xps}, \ref{sec:raman}). 

In the examined lamellae (Figure~\ref{fgr:HRTEM}a,c,e,g), the coatings are locally continuous on Au, with no obvious delamination or interfacial voids. This is evidence of the imaged contact geometry, not a measurement of adhesion strength or long-term mechanical reliability. The carbon/Pt protection layers were added for FIB preparation and are not functional device electrodes. The film thickness is $1.75\pm0.50$~$\mu$m for the PB-like film and $215\pm100$~nm for the PW-like film, where the $\pm$ values give the roughness of the faceted surfaces; independent thickness distributions across the potential series are not available.

Contrast changes reveal differently oriented crystallites. The indexed regions include [001], [011], and [111] orientations, but their limited sampling does not establish a quantitative whole-film texture. HRTEM fringes and Fourier transforms are consistent with local cubic frameworks (Figure~\ref{fgr:HRTEM}b,d,f,h). For example, the [001] PB region contains reflections indexed as (200), (020), and (220). Lattice parameters from multiple selected regions were reported as $(10.31\pm0.10)$\,\AA\ for PB and $(10.36\pm0.10)$\,\AA\ for PW. Reciprocal spacings were extracted from distances between reflections and the zero beam in the Fourier transforms, using
\[
d_{hkl}=\frac{a}{\sqrt{h^2+k^2+l^2}}.
\]
The uncertainties describe the variation between measurements; magnification calibration, Fourier-transform precision, and local strain further limit the absolute values. The two values overlap within their uncertainties, so the lattice parameter cannot discriminate the two redox and compositional states. This is consistent with the literature: reported PB/PW lattice-parameter changes in K$^{+}$ media are only about 1--2\%. In situ XRD during reduction in KCl showed an increase of 0.082~\AA\ from 10.16~\AA\ with the cubic cell retained,\cite{Ikeshoji1988} whereas ex situ XRD of films electrodeposited from the same bath composition (to 50~mC) showed a decrease from about 10.23~\AA\ (PB) to 10.06~\AA\ (PW), with higher values for thin films grown at $-0.10$~V.\cite{Baggio2019} Such changes are comparable to the uncertainty of the present measurement. No XRD phase fraction or cubic/rhombohedral transformation was measured in this study. However, a deeper structural analysis using x-ray methods are in scope of a future investigation.

STEM--EELS spectra resolve the characteristic C, K, N, O, and Fe core-loss features, including Fe~L$_3$ and L$_2$ edges (Figure~\ref{fgr:EELS}b,e). Because the C-K and K-L$_{2,3}$ edges partly overlap, C and K are shown as a combined map rather than two independently quantified distributions. The Fe and N maps are laterally homogeneous within the examined regions of both the films (Figure~\ref{fgr:EELS}c,f). A localized carbon-rich region in PB does not track the Fe/N framework distribution. Weak oxygen contrast and apparent light-element signals in Au can include background-subtraction, beam-delocalization, and through-thickness projection contributions.\cite{Egerton2011} These possibilities limit an inference of either true interdiffusion or its complete absence.

\begin{figure*}[!t]
 \centering
    \includegraphics[width=0.85\linewidth, keepaspectratio]{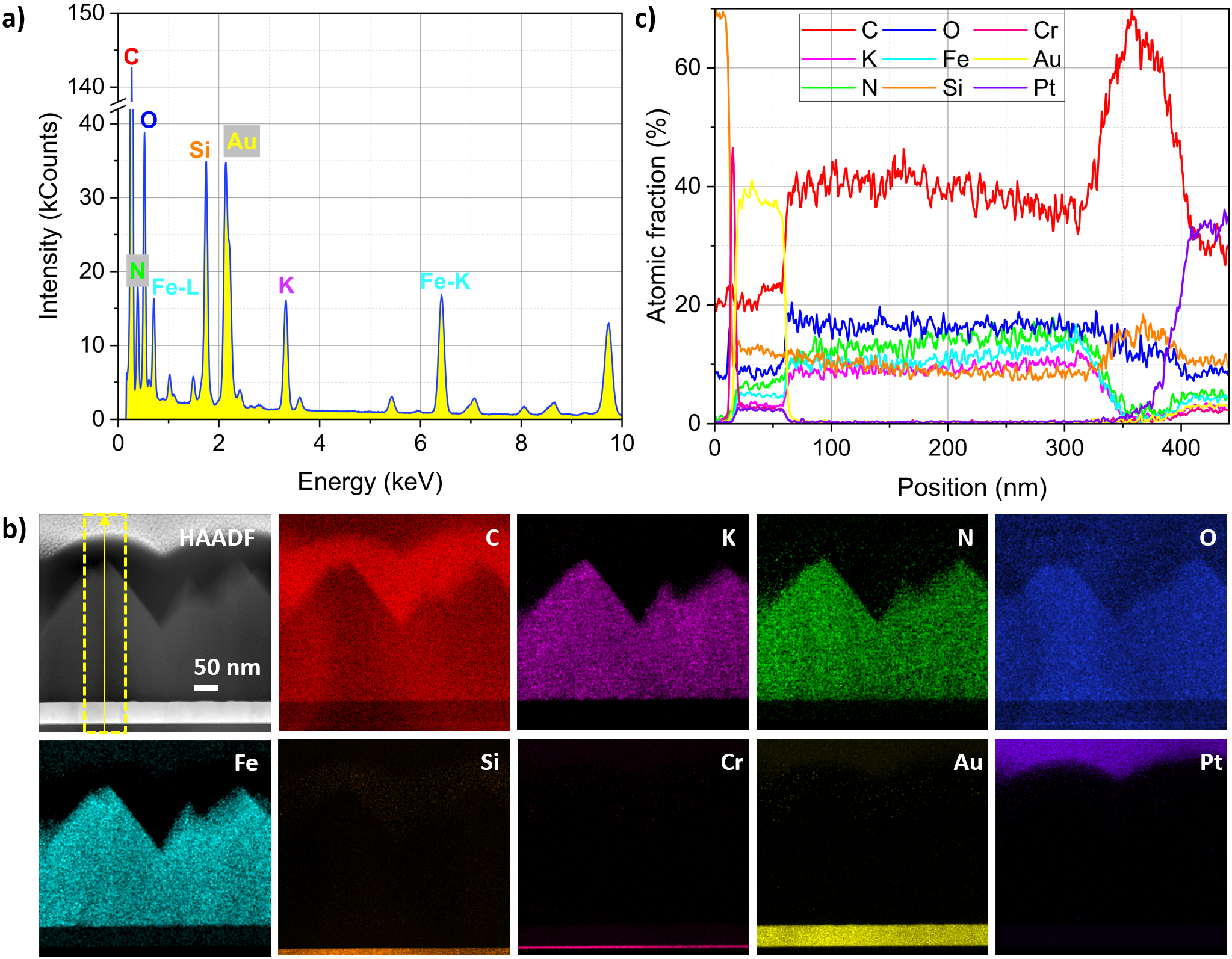}
   \caption{STEM--EDS of the PW-like film ($-0.10$~V), shown as in Fig.~8. (a) Spectrum from the analyzed region. (b) HAADF-STEM image and elemental maps. (c) Line profile from the Si substrate to the top of the lamella along the yellow-marked region in (b).}
 \label{fgr:EDS_PW}
\end{figure*}

\begin{table}[h]
\centering
\caption{STEM--EDS-derived K/Fe ratio and K--Fe-normalized composition of the PB-like ($+0.30$~V) and PW-like ($-0.10$~V) films. Values are reported as mean $\pm$ standard deviation of the point-by-point values along one line profile per film, drawn parallel to the Au interface within the film (see Section 2.2).}
\label{tab:stemeds-ratios}
 
\small
\renewcommand{\arraystretch}{1.15}
\setlength{\tabcolsep}{4pt}
 
\begin{tabular*}{\linewidth}{@{}r@{\extracolsep{\fill}}rrr@{}}
\toprule
\shortstack{$E_{\mathrm{dep}}$\\(V)}
& K/Fe
& \shortstack{$100\,\mathrm{K}/(\mathrm{K}+\mathrm{Fe})$\\(\%)}
& \shortstack{$100\,\mathrm{Fe}/(\mathrm{K}+\mathrm{Fe})$\\(\%)} \\
\midrule
 
$+0.30$
& $0.530 \pm 0.088$
& $34.42 \pm 3.87$
& $65.58 \pm 3.87$ \\
 
$-0.10$
& $0.853 \pm 0.046$
& $45.99 \pm 1.33$
& $54.01 \pm 1.33$ \\
 
\bottomrule
\end{tabular*}
 
\par\vspace{4pt}
\begin{minipage}{\linewidth}
\footnotesize
K/Fe and both normalized fractions were calculated separately for each profile point before determining their means and standard deviations. Light elements (C, N, O) are not included because their quantification is affected by the FIB-related protection layers and carbon redeposition.
\end{minipage}
\end{table}

STEM--EDS maps and line profiles (Figures~\ref{fgr:EDS_PB} and \ref{fgr:EDS_PW}) show C, K, N, O and Fe in the films, Au and Cr in the electrode and adhesion layers, and Pt and part of the carbon signal in the protection layers. Fe, N and K are present throughout both films without pronounced segregation at the displayed resolution. The profiles separate the film, support and protection layers, supporting locally defined interfaces without establishing an atomically abrupt boundary.

Unlike the EELS data, both STEM--EDS datasets were acquired on the same instrument and quantified with the same procedure. The K/Fe ratio is taken within each spectrum and is therefore independent of absolute signal intensity, and the background-subtracted quantification accounts for the different energies of the K K$\alpha$ (3.3~keV) and Fe K$\alpha$ (6.4~keV) lines. Within these limits, the ratio allows a direct compositional comparison of the two films. Along one line profile parallel to the Au interface within each film (2.67~$\mu$m with 468 points for PB-like; 468~nm with 660 points for PW-like), K/(K+Fe) is $34.4\pm3.9$\% for the PB-like and $46.0\pm1.3$\% for the PW-like film (Table~2). The uncertainties are the standard deviation along each profile and describe local variation within one lamella, not reproducibility between independently deposited films. The cumulative electron dose was about $1.6\times10^{3}$~e$^{-}$\,\AA$^{-2}$ for the PB-like and $7.6\times10^{4}$~e$^{-}$\,\AA$^{-2}$ for the PW-like film. Beam-induced K migration would lower the measured K fraction, so the much higher dose on the PW-like film can only have reduced, not created, its higher K content relative to the PB-like film. Beam-induced K loss could not be assessed directly, because individual frames were not stored; the measured K fractions may therefore be lower bounds, particularly for the higher-dose PW-like film, and any such loss can only have reduced the observed PB/PW contrast.

The lateral K distribution also differs between the films. In the PW-like film, K/(K+Fe) averaged over the first, middle and last third of the profile is 46.0, 46.1 and 45.9\%, respectively. In the PB-like film, the corresponding values are 29.7, 35.7 and 37.9\%, indicating lateral compositional heterogeneity on the micrometer scale. The profiles across the film thickness (Fig.~\ref{fgr:EDS_PB}c and \ref{fgr:EDS_PW}c) show no pronounced K enrichment at the film surface or the Au interface at their resolution. The carbon fraction in the PW-like film exceeds the framework stoichiometry, which we attribute to carbon redeposited on the thin lamella during FIB preparation. C, N and O are therefore treated qualitatively, and only the K/Fe ratio is interpreted quantitatively.
 
The STEM--EDS ratios can be compared with the SEM--EDS ratios (Fig.~\ref{fgr:EDS_results}h; Fe/K of 3.09 and 1.38, i.e.\ K/(K+Fe) of 24.4\% and 42.0\%), the XPS values (Table~\ref{tab:xps-ratios-total}) and the idealized limiting compositions, 33.3\% for KFe[Fe(CN)$_{6}$] and 50\% for K$_{2}$Fe[Fe(CN)$_{6}$]. XPS gives similar K-rich surface compositions for both films, 55.5\% at $+0.30$~V and 56.9\% at $-0.10$~V, above even the idealized PW value. The cross-sectional STEM--EDS values are lower for both films and, unlike XPS, clearly separate them. The difference between the two methods is largest for the PB-like film (34.4\% vs.\ 55.5\%), indicating that the K-rich composition within the few-nanometre XPS sampling depth is not representative of the film interior; surface enrichment, residual K-containing species from the KCl-containing bath, or differences in quantification between the methods are possible contributions. For the PW-like film, the STEM--EDS value of 46.0\% lies between the SEM--EDS (42.0\%) and XPS (56.9\%) values, indicating a moderately K-deficient PW-like film. For the PB-like film, the STEM--EDS value (34.4\%) is close to the idealized soluble-PB composition and lies between the SEM--EDS (24.4\%) and XPS (55.5\%) values. The cross-sectional K/(K+Fe) ratio is thus the clearest difference between the PB-like and PW-like films, whereas neither the lattice parameter nor the surface K fraction from XPS discriminates them. Because stoichiometry was not refined and vacancy and water contents are unknown, these ratios are compositional observations rather than formula assignments.
 
Overall, the TEM results show PB-like and PW-like films with a locally cubic framework and continuous interfaces to the Au electrode, which differ mainly in their cross-sectional K content, consistent with the more reduced Fe state indicated by XPS.

\subsection{Synthesis--structure relationships and evidential limits}\label{sec:synthesis-structure}
The joint observations support a K-richer, more reduced PW-like environment at lower potentials and K-poorer PB-like environments at higher potentials, with different local morphologies but overlapping cubic lattice parameters. ``PB'' and ``PW'' are used as shorthand for these combined spectroscopic/compositional assignments, not as quantitative declarations of phase purity. For an ideal vacancy-free framework, a useful redox/charge-compensation relation is
\[
\mathrm{KFe^{III}[Fe^{II}(CN)_6]+K^++e^-}
\rightleftharpoons\mathrm{K_2Fe^{II}[Fe^{II}(CN)_6]}.
\]
This expresses a possible framework reduction step, not the complete pathway by which FeCl$_3$ and ferricyanide form the deposited solid. Solution-species reduction, nucleation kinetics, and transport may also contribute. The common acidic bath affects precursor speciation, but local pH was not measured and cannot be invoked as an established cause of the crossover.

The room-temperature, single-deposition route provides films directly supported on Au and allows different initial K/redox states to be compared without a slurry binder. However, equal growth charge does not isolate composition from differences in loading or morphology, and selected local maps do not establish full-area uniformity.

\section{Single-Step Electrodeposited PBAs: Applications and Outlook}\label{sec:applications-outlook}

\subsection{Alkali-ion storage and thin-film capacitive electrodes}
Single-step electrodeposition offers a straightforward route to PB and PW thin-film electrodes, forming the coating directly on a conducting support without polymer binders or added conductive powders. Their open cyanide-bridged framework can accommodate reversible alkali-ion insertion and extraction coupled to metal-centre redox reactions, providing the basis for battery applications.\cite{Wang2018_PBA_RechargeableBatteries,Xiao2024_PBA_SIB_Review} In the present films, the variation in K content and chemical environment (Sections~\ref{sec:xps} and \ref{sec:raman}) provides starting conditions for comparing K-ion storage across deposition potentials. Measurements of reversible capacity, operating potential, and cycling stability would establish suitable cell configurations, with alkali-inventory retention and framework stability as important evaluation criteria.\cite{Nielsen2024_PW_FullCells_OperandoXRD,Zhang2025_PrussianBlueCapacityFading}

Beyond batteries, PB/PW redox chemistry also motivates thin-film supercapacitors and micro-supercapacitors. Published demonstrations using PB/reduced graphene oxide and mixed Fe--Co hexacyanoferrate electrodes illustrate the energy-storage potential of this material family.\cite{Avila2024PB,Mathi2025_FeCoHexacyanoferrateSupercapacitor} Here, direct growth on Au and the film morphology characterized in Sections~\ref{sec:morphology} and \ref{sec:tem} offer a platform for examining how electrode preparation influences charging kinetics and areal charge storage. Rate-dependent voltammetry and galvanostatic cycling would assess the contribution of capacitive processes, while full-cell measurements would establish usable capacitance, energy density, power density, and durability. These results would clarify the suitability of the coatings for compact electronic power sources.

\subsection{Electrochemical sensor applications}\label{sec:outlook-sensing}
PB/PW films offer a promising starting point for electrochemical sensors and biosensors. PB-modified electrodes exhibit well-established ``artificial peroxidase'' activity towards hydrogen peroxide reduction.\cite{Karyakin1999_ArtificialPeroxidase} Peroxide provides an analytical readout when an appropriate oxidase generates H$_2$O$_2$ from the target substrate. Published examples include glucose and lactate biosensors operating at 0.00 and $-0.05$~V versus Ag/AgCl, respectively, and cholesterol biosensors incorporating printed PB nanoparticles.\cite{Lin2015_GlucosePDA,Garjonyte2001_LactatePB,Cinti2015_CholesterolPB} These low operating potentials provide a basis for limiting interference from readily oxidized compounds. Additional selectivity can be introduced through functional layers: optimized PB/poly(toluidine blue) glucose biosensors have suppressed interference from ascorbic acid, uric acid, and acetaminophen.\cite{Zhang2004_GlucoseMultilayerPB}

In the present work, direct electrodeposition forms supported coatings on Au without an added deposition binder. This geometry, together with the observed variations in morphology and composition, provides a platform for investigating how deposition conditions influence sensing performance. Subsequent functionalization could introduce enzyme recognition layers or protective coatings; polydopamine-coated PB/Au electrodes, for example, have demonstrated improved stability at neutral pH and rejection of selected interferents.\cite{Gao2016_ProtectivePDA} Printed cholesterol sensors and wearable glucose/lactate arrays further demonstrate the integration potential of PB-based transduction.\cite{Cinti2015_CholesterolPB,Gao2016_WearableSweat} These precedents motivate developing the present coatings towards compact analytical devices, with sensitivity, linear range, detection limits, mechanical durability, and selectivity established through measurements in the intended sample matrix. Single-step electrodeposition thus offers a compelling route for sensor fabrication, streamlining the direct integration of PB/PW coatings onto conductive supports and establishing a versatile platform for subsequent enzymatic or protective functionalization. Crucially, this streamlined, binder-free methodology overcomes the reproducibility and scalability bottlenecks of traditional multi-step coating techniques, making it an indispensable approach for the development of reliable, mass-producible electrochemical biosensors.

\subsection{Resistive switching and memristive devices}\label{sec:outlook-memory}
The K-richer, more reduced PW-like films and the K-poorer PB-like films provide contrasting starting states for investigating how composition and Fe chemical environment influence electrically induced resistance changes (Sections~\ref{sec:xps} and \ref{sec:raman}, Table~\ref{tab:xps-ratios-total}, and Figures~\ref{fgr:XPS results} and \ref{fig:raman_pb_pw}). The locally resolved Au/film interfaces (Section~\ref{sec:tem}, Figures~\ref{fgr:HRTEM}, \ref{fgr:EDS_PB}, and \ref{fgr:EDS_PW}) provide a supported-film geometry that could be completed with a top electrode. These observations define structural and compositional variables for comparing device responses, without establishing the electrical properties of the present specimens.

Published Au/PB/Ag stacks have exhibited bipolar resistive switching with a current contrast of at least three orders of magnitude over 200 tested cycles.\cite{Avila2020_ResistiveSwitching_PrussianBlue} PW-based films have also shown resistive switching, including systems with unipolar behaviour.\cite{Faita2022_PrussianWhite_ResistiveSwitching,Avila2022_UnipolarRS_PW_ThinFilms} Related PB studies have examined cycle-to-cycle variability,\cite{Avila2025} conductance quantization,\cite{Cantudo2026} and nanoscale conductance modulation interpreted in terms of K$^+$ redistribution coupled to Fe redox changes.\cite{Avila2026,Avila2024Perylene}. These published results motivate composition- and contact-controlled tests; they do not determine the switching polarity, resistance contrast, or operating voltage of the present films.

Device geometry and interfacial chemistry are additional factors governing resistive switching. Torres-Cavanillas et al. integrated large-area two-dimensional Prussian blue into lateral Ag/2D-PB/Au nonvolatile memristors, attributing their bipolar switching to an electrochemical metallization mechanism involving Ag$^{+}$ migration through the PB framework and the reversible formation and rupture of conductive Ag filaments.\cite{TorresCavanillas2026} The continuous 2D morphology and lateral transport geometry differ from those of the granular electrodeposited coatings characterized here, whose switching mechanism requires independent experimental investigation. Ide et al. used Ru-complex primer layers between ITO and PB to rectify electron transfer and gate K$^+$ storage/release, supported by electrochemical and electrogravimetric measurements.\cite{Ide2021} Such results motivate tests of interfacial control in PB-based devices; electrochemical ion gating alone does not demonstrate retained resistance states. 

In a distinct printed architecture, He et al. reported a nonvolatile Ag/high-entropy-PBA/ITO memristor with mean switching voltages of approximately $V_{\mathrm{SET}}=-0.26$\,V and $V_{\mathrm{RESET}}=+0.36$\,V, an $R_{\mathrm{OFF}}/R_{\mathrm{ON}}$ ratio of approximately $10^{4}$, and reported SET and RESET powers of 26 and 8\,$\mu$W, respectively. The device maintained switching over 800 voltage-sweep cycles. Its resistance changes were attributed to Na$^+$ extraction/insertion coupled to changes in electronic conductivity.\cite{He2025_HEPBA_Memristor} These figures describe a Na-containing, multimetal, printed device and provide external benchmarks for comparison, not measured or predicted performance of the present K-containing, Fe-only films. Framework-cation redistribution must be tested alongside electrode-metal migration, filament formation, contact-barrier changes, and Joule heating; static elemental maps cannot discriminate these mechanisms.

The room-temperature, single-step electrodeposition process forms the active material directly on a conducting support and permits systematic adjustment of deposition conditions. In our previous work, C-AFM measurements demonstrated reversible resistive switching localized within sub-100 nm regions of electrodeposited PB/PW films, providing evidence for an intrinsic switching mechanism involving K$^{+}$ redistribution coupled to Fe$^{2+}$/Fe$^{3+}$ redox reconfiguration within the framework.\cite{Avila2026} Similarly, He et al. attributed switching in printed high-entropy PBA memristors to Na$^{+}$ extraction/insertion and the resulting modulation of framework conductivity, supporting an intrinsic ion-mediated response rather than switching governed by Ag-filament formation.\cite{He2025_HEPBA_Memristor}
These findings support the development of solution-processed PBA memristors, while translating the local material response into integrated devices requires systematic assessment of endurance, retention, variability, and fabrication compatibility.\cite{Pazos2024} 
\subsection{Catalytic films and precursors}\label{sec:outlook-catalysis}
The broader catalytic and photocatalytic applications of PBAs have been reviewed across different compositions and reaction systems.\cite{Concina2024_PBA_CatalystsReview} Section~\ref{sec:morphology} documents faceted surfaces; Sections~\ref{sec:xps} and \ref{sec:raman} establish cyanoferrate chemical environments; Section~\ref{sec:tem} resolves local coating/support contact. These are relevant preparation variables for catalytic comparisons, but do not quantify active-site density or establish Oxygen evolution reaction/Oxygen reduction reaction (OER/ORR) activity. In particular, the undoped Fe-only films cannot inherit the behaviour of Co-, Ni-, or Mn-containing catalysts. Primary research on Ni-doped Co--Fe PBAs identifies reconstructed oxyhydroxide sites during OER,\cite{Hou2021_Reconstruction} whereas an ORR study converts PB with an additional nitrogen source into FeN$_x$/Fe-containing carbon nanotube material by pyrolysis.\cite{Zhao2024_PBDerivedORR} The latter is a derived catalyst, not an intact electrodeposited PB/PW film; both activation and carbonization add steps beyond single-step film growth.

The experimentally mapped Fe/C/N distribution can motivate testing whether a supported precursor gives a controlled spatial distribution after conversion, but neither framework retention nor contact stability through activation is known. No high-spin Fe(III)/Fe(IV) catalytic cycle or catalytic phase-boundary enhancement can be inferred from the present XPS/HRTEM data. Appropriate validation includes substrate and composition controls, active-area/metal-loading normalization, product-selective Faradaic efficiency, dissolution measurements, and operando or post-reaction spectroscopy/diffraction to identify the actual active phase. Claims about water-splitting or oxygen-reduction performance should await those measurements.

\section{Conclusions}\label{sec:conclusions}
Room-temperature, single-step electrodeposition produced directly supported PB and PW films with potential-dependent composition and morphology. Lower deposition potentials favoured K-richer, more reduced PW-like deposits, whereas higher potentials yielded K-poorer PB-like films. Spectroscopy and electron microscopy confirmed cyanoferrate framework features with representative PB-like and PW-like regions exhibiting cubic crystallinity. In addition, laterally homogeneous elemental distributions within the representative PB and PW films and well-defined interfaces with the Au/Cr/Si substrate stack were found, providing nanoscale verification of the structural and chemical integrity of the deposited layers. Overall, this study establishes a direct correlation between the electrodeposition potential and the resulting morphology, composition, and crystal structure of electrodeposited PB and PW thin films. Direct growth on a conducting support without a slurry binder offers a useful advantage for integrating these films into functional architectures. Access to contrasting K/redox states within a common deposition route further supports application-specific optimization. K-rich and K-poor coatings provide complementary starting points for alkali-ion storage studies, while the PB/PW chemical contrast motivates peroxide sensing, electrochromism, and resistive switching. The combined findings provide a strong experimental foundation for developing supported cyanoferrate films, with targeted electrochemical, optical, and electrical measurements defining the next step towards improved application performance.

\section*{Author contributions}

\textbf{Conceptualization:} M. P., L.B.A.; 
\textbf{Methodology:} M. P., L. O. G., M. F. K., D. P., F. R.;
\textbf{Investigation:} M. P., M. F. K. L.B.A; 
\textbf{Validation:} M. P., M. F. K.;
\textbf{Data curation:} M. P., M. F. K. L.B.A;
\textbf{Formal analysis:} M. P., L.B.A, M. F. K.; 
\textbf{Visualization:} M. P., M. F. K.;
\textbf{Supervision:} C. K. M., D. W.; A. L.;
\textbf{Writing -- original draft:} M. P., M. F. K.;
\textbf{Writing -- review \& editing:} M. P., M. F. K., C. K. M., D. W., A. L.;

\section*{Conflicts of interest}
There are no conflicts to declare.

\section*{Data availability}
The reported characterization results are presented in the article and accompanying supporting information. Requests for original spectra, images, fitting files, and the complete replicate results should be directed to the corresponding author.

\section*{Acknowledgements}
This research was funded by the European Social Fund (ESF) and the Free State of Saxony within the framework of the Landesinnovationspromotion program (application number 100670500), as well as by the Deutsche Forschungsgemeinschaft (DFG, German Research Foundation) -- Project number 531524052. XPS analyses were performed by Pierre Eloy (UCLouvain), whose assistance is gratefully acknowledged. We acknowledge the use of the facilities in the Dresden Center for Nanoanalysis (DCN) at the Technische Universität Dresden.

\balance

\bibliography{rsc} %
\bibliographystyle{rsc} %

\end{document}


\begin{center}
{\large\bfseries Supplementary Information}\\[2pt]
RSC Advances\\[10pt]
{\Large\bfseries Prussian blue and Prussian white thin-film electrodes: Single-step electrodeposition for anode and cathode applications\par}
\vspace{10pt}
Michael Pohlitz$^{a}$, Mohammed Fayis Kalady$^{b}$, Larissa de O.\ Garcia$^{a}$, Lindiomar Borges de Avila Jr$^{c}$, Darius Pohl$^{d}$, Falk R\"oder$^{b}$, Flavio Abreu Araujo$^{c}$, Axel Lubk$^{b,e}$, Daniel Wolf$^{b}$ and Christian K.\ M\"uller$^{a,*}$
\end{center}

{\small
$^{a}$\,Faculty of Physical Engineering/Computer Sciences, University of Applied Sciences Zwickau, 08056 Zwickau, Germany.\\
$^{b}$\,Leibniz Institute for Solid State and Materials Research Dresden, 01069 Dresden, Germany.\\
$^{c}$\,Institute of Condensed Matter and Nanosciences (ICMN), Universit\'e catholique de Louvain (UCLouvain), Louvain-la Neuve, B-1348, Belgium.\\
$^{d}$\,Dresden Center for Nanoanalysis, Center for Advancing Electronics Dresden, TU Dresden, 01069 Dresden, Germany.\\
$^{e}$\,Institute of Solid State and Materials Physics, TU Dresden, 01069 Dresden, Germany.\\ 
$^{*}$\,Corresponding author: \href{mailto:Christian.Mueller.1@whz.de}{Christian.Mueller.1@whz.de}
}

\tableofcontents

This document presents the XPS survey spectra, archived surface atomic concentrations, fitted-component labels, binding energies, and full widths at half maximum (FWHM). The accompanying CSV files preserve the tabulated numerical values in a machine-readable form.

\clearpage
\section{XPS survey spectra}

\begin{figure}[h]
\centering
\includegraphics[width=0.85\linewidth]{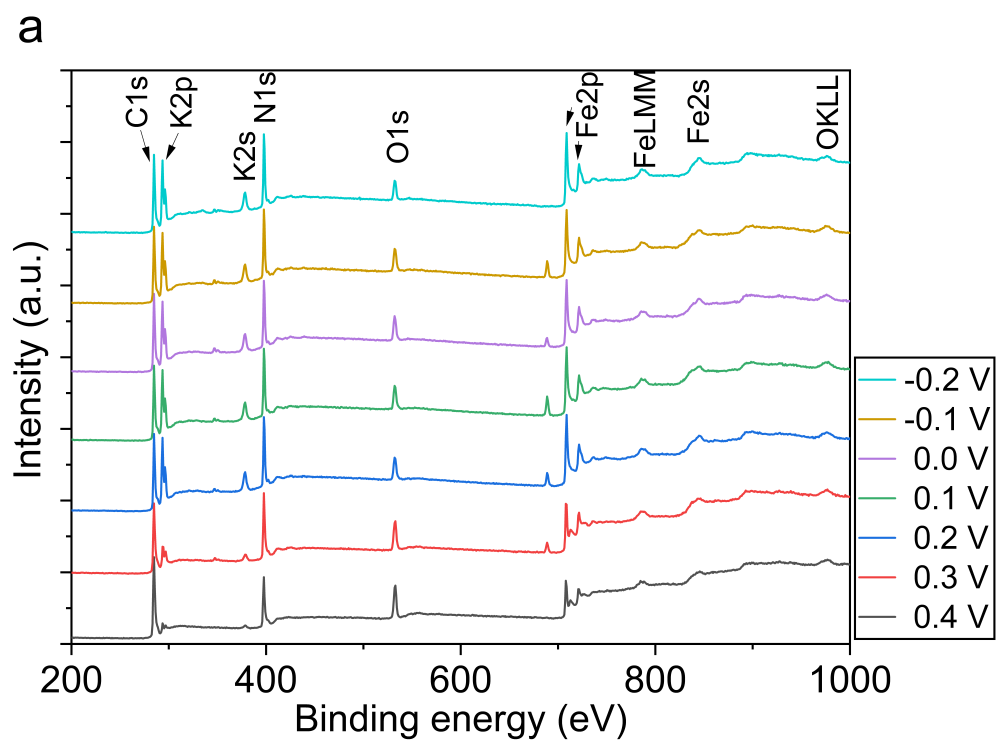}
\caption{XPS survey spectra of PB/PW-family films deposited at $-0.20$, $-0.10$, 0.00, $+0.10$, $+0.20$, $+0.30$ and $+0.40$\,V versus SCE, displayed from top to bottom as indicated by the colour legend. The displayed binding-energy range is approximately 200--1000\,eV. Curves are vertically displaced for clarity and intensity is shown in arbitrary units. The original labels identify C\,1s, K\,2p, K\,2s, N\,1s, O\,1s, Fe\,2p and Fe\,2s photoelectron features, together with Fe\,LMM and O\,KLL Auger features. Relative peak heights are qualitative and are not a substitute for sensitivity-corrected atomic ratios.}
\label{fig:S1}
\end{figure}

Fig.~\ref{fig:S1} establishes the presence of the main detected constituents across the deposition-potential series. The C\,1s signal near 285\,eV and N\,1s signal near 398\,eV accompany the Fe\,2p region near 708--724\,eV, consistent with the elemental constituents of cyanoferrate films. K is identified by the neighbouring K\,2p feature near 294\,eV and the K\,2s feature near 378\,eV. The O\,1s signal near 532\,eV shows that oxygen-containing species also contribute within the XPS sampling depth. The survey alone does not distinguish framework-associated water from adsorbed water, surface oxygen-containing species, or other oxygen contributions.

The relative amplitudes of the labelled features vary across the series, and the K features are visually less prominent in the two highest-potential traces. However, the spectra are offset, and the original intensity scaling and normalization are not documented. Element-dependent sensitivity and background contributions also affect the plotted peak heights.\cite{Shard2020} Consequently, the displayed traces support qualitative elemental identification but do not independently determine the revised K/Fe statistics, exact bulk formulas, or PB/PW phase fractions. Binding energies and fitted widths must be assessed from the tabulated channels and the original narrow-scan fits, rather than read precisely from this survey overview.

\clearpage
\section{Surface composition and fitted components}

Table~\ref{tab:S2} reports the atomic concentrations of Fe, F, O, N, Ca, K, C and Au in the archived analyses. N\,1s is represented by three fitted components (N1--N3) and C\,1s by four (C1--C4); the separate N and C total columns provide the reported elemental concentrations. Component labels are defined in Table~\ref{tab:S1} and used in the same order in the binding-energy and FWHM tables. A total column is not an additional spectral component, so adding it to its component columns would double-count that element. Totals are transcribed as reported instead of reconstructed from rounded component values.

Within this archive, Fe concentrations are 6.9--7.9\,at.\% from $-0.20$ to $+0.20$\,V, decreasing to 5.9 and 4.6\,at.\% in the single analyses at $+0.30$ and $+0.40$\,V. K is 8.9--9.7\,at.\% over the lower-potential range and is 3.5 and 1.2\,at.\% in those two higher-potential archived analyses. These are properties of the earlier export, not the updated four-analysis means in manuscript Table~1. They should not be combined with the revised
K--Fe statistics to construct a new composition or a new uncertainty estimate.

The reported total C fraction is 44.9--47.6\,at.\% up to $+0.20$\,V, increasing to 49.5 and 59.5\,at.\% at $+0.30$ and $+0.40$\,V. Total N is 25.5--28.3\,at.\% up to $+0.20$\,V, followed by 27.7 and 23.0\,at.\%, respectively. O increases from a lower-potential range of 6.4--8.8\,at.\% to 10.9 and 11.8\,at.\%. Thus the C and O fractions rise as the Fe fraction falls in the archived high-potential analyses. Since atomic fractions are normalized quantities, these changes do not by themselves measure an absolute accumulation of C or O. Additional surface species, differences in
near-surface composition, and attenuation effects remain possible contributions.\cite{Shard2020}

\begin{table}[h]
\centering
\caption{Meaning of the N\,1s and C\,1s component labels used consistently
in the supplementary tables.}
\label{tab:S1}
\small
\begin{tabular}{@{}llll@{}}
\toprule
Component & BE range (eV) & Source attribution & Interpretation of the label \\
\midrule
N1 & 402.0--402.2 & No attribution supplied & Highest-binding-energy fitted N component \\
N2 & 399.1--399.8 & No attribution supplied & Intermediate-binding-energy fitted N component \\
N3 & 397.5--397.9 & ``Nitride'' & Source label; not evidence of a separate nitride phase \\
C1 & 288.8--288.9 & O=C--O & Source attribution retained \\
C2 & 287.6 & C=O, O--C--O & Source attribution retained \\
C3 & 286.1--286.2 & C--(O,N) & Source attribution retained \\
C4 & 284.6--284.8 & C--(C,H) & Source attribution retained \\
\bottomrule
\end{tabular}
\end{table}

The source names for N and C are retained for traceability. In particular, N3 is labelled ``Nitride'' in the export, but that software attribution is not sufficient to identify a nitride phase in a cyanoferrate film. N1 and N2 have no supplied attribution. Similarly, the C1--C4 labels are not a validated separation of framework cyanide carbon from adventitious or oxygenated surface carbon. Component-specific assignments require the corresponding spectra, fitting model, and reference evidence.\cite{Major2020}

F and Ca are identified as survey estimates. Where reported, F is 1.8--3.9\,at.\% and Ca is 0.3--0.6\,at.\%. These entries do not establish a unique contamination source or intentional incorporation into the framework. Au values range from 0.02 to 1.90\,at.\% where listed. Some are below the approximately 0.1\,at.\% detection limit stated in the source and are retained as reported, rather than interpreted as secure quantitative detections. The Au channel is labelled Au[0] and 4f$_{5/2}$ in the export; its orbital designation requires verification against the original Au doublet.

Counting N and C once, the sums of the listed atomic percentages range from 99.80 to 100.15\,at.\%. Near-complete closure checks transcription and normalization consistency; it does not establish phase purity, eliminate unmeasured species such as H, or validate each individual peak assignment.

\begin{table}[htbp]
\centering
\caption{Surface atomic concentrations (at.\%) and fitted N\,1s and C\,1s contributions in all 13 archived analyses.}
\label{tab:S2}
\scriptsize
\setlength{\tabcolsep}{3.5pt}
\begin{tabular}{@{}llrrrrrrrrrrrrrrr@{}}
\toprule
 & $E_{\mathrm{dep}}$ & Fe & F$^{a}$ & O & \multicolumn{4}{c}{N\,1s} & Ca$^{a}$ & K & \multicolumn{5}{c}{C\,1s} & Au$^{b}$ \\
\cmidrule(lr){6-9}\cmidrule(lr){12-16}
ID & (V vs.\ SCE) & 2p$_{3/2}$ & 1s & 1s & N1 & N2 & N3 & Total & 2p$_{3/2}$ & 2p & C1 & C2 & C3 & C4 & Total & 4f$_{5/2}$ \\
\midrule
10:1 & $-0.20$ & 7.9 & --- & 6.4 & 1.3 & 0.9 & 26.1 & 28.3 & 0.4 & 9.6 & 2.0 & 1.9 & 3.6 & 39.7 & 47.2 & 0.35 \\
11:1 & $-0.20$ & 7.5 & --- & 7.3 & 1.2 & 0.7 & 25.4 & 27.3 & 0.4 & 9.7 & 2.1 & 2.1 & 3.6 & 39.6 & 47.4 & 0.34 \\
12:1 & $-0.20$ & 7.4 & --- & 7.5 & 1.1 & 0.7 & 25.4 & 27.2 & 0.5 & 9.6 & 2.1 & 2.0 & 3.7 & 39.7 & 47.6 & 0.22 \\
13:1 & $-0.20$ & 7.4 & --- & 6.5 & 1.3 & 0.8 & 25.0 & 27.1 & 0.3 & 9.2 & 1.9 & 1.9 & 3.8 & 39.8 & 47.4 & 1.90 \\
1:1 & $-0.10$ & 7.0 & 3.3 & 8.2 & 1.2 & 0.7 & 24.1 & 26.0 & 0.4 & 9.1 & 2.1 & 1.9 & 3.6 & 38.3 & 45.8 & 0.16 \\
2:1 & $-0.10$ & 7.0 & 2.1 & 8.8 & 1.0 & 0.5 & 23.9 & 25.5 & 0.5 & 9.4 & 2.2 & 2.0 & 3.6 & 38.9 & 46.7 & 0.02 \\
3:1 & $-0.10$ & 7.0 & 2.5 & 8.7 & 1.0 & 0.5 & 24.2 & 25.7 & 0.5 & 9.5 & 2.2 & 1.8 & 3.4 & 38.6 & 46.1 & 0.03 \\
4:1 & $-0.10$ & 6.9 & 1.9 & 8.4 & 1.1 & 0.8 & 23.6 & 25.6 & 0.4 & 8.9 & 2.2 & 2.0 & 4.0 & 39.3 & 47.5 & 0.51 \\
5:1 & $0.00$ & 7.0 & 1.8 & 8.0 & 1.1 & 0.5 & 24.4 & 26.0 & 0.6 & 9.3 & 2.2 & 1.8 & 3.6 & 39.5 & 47.1 & 0.03 \\
6:1 & $+0.10$ & 7.2 & 3.9 & 7.8 & 1.2 & 0.7 & 24.6 & 26.5 & 0.4 & 9.3 & 2.0 & 1.8 & 3.4 & 37.7 & 44.9 & --- \\
7:1 & $+0.20$ & 7.5 & 2.6 & 7.6 & 1.1 & 0.6 & 25.2 & 26.8 & 0.4 & 9.5 & 2.0 & 1.9 & 3.4 & 38.2 & 45.6 & --- \\
8:1 & $+0.30$ & 5.9 & 2.2 & 10.9 & 1.2 & 3.9 & 22.6 & 27.7 & 0.4 & 3.5 & 2.1 & 2.7 & 6.5 & 38.2 & 49.5 & --- \\
9:1 & $+0.40$ & 4.6 & --- & 11.8 & 1.6 & 4.1 & 17.3 & 23.0 & --- & 1.2 & 2.1 & 2.8 & 7.2 & 47.4 & 59.5 & --- \\
\bottomrule
\end{tabular}

\vspace{4pt}
\begin{minipage}{\linewidth}
\footnotesize N and C totals are the reported elemental totals and are counted once in elemental sums. Source precision is retained, including the two-decimal Au concentrations. The analysis IDs match Tables~\ref{tab:S3} and \ref{tab:S4}; the full set is also provided in \texttt{anc/data/atomic\_concentrations.csv}, where empty cells correspond to dashes here.

\footnotesize $^{a}$\,F and Ca are identified as survey estimates in the source. A dash denotes a blank source entry, not zero or a demonstrated absence of the element. $^{b}$\,The Au orbital label is reproduced as 4f$_{5/2}$ from the export; the label requires verification against the original Au doublet. N3 and the C-component names are source labels, as explained in Table~\ref{tab:S1}.
\end{minipage}
\end{table}

\section{Binding energies}

Table~\ref{tab:S3} lists the reported binding energies in eV for the same analysis IDs and component ordering as the concentration table. A binding energy locates a fitted peak on the energy scale; it is not a component area or a direct measure of the amount of that chemical state. Component assignments depend on energy referencing, spectral line shapes, the background, and consistency with chemically plausible fits.\cite{Major2020}

The exported Fe\,2p$_{3/2}$ channel remains at 708.7\,eV from $-0.20$ through $+0.20$\,V, then appears at 708.5\,eV at $+0.30$\,V and 708.3\,eV at $+0.40$\,V. This channel is not the complete Fe\,2p multiplet decomposition shown separately in the manuscript. Its position alone does not provide an Fe(II)/Fe(III) area ratio, and the small shifts cannot be assigned uniquely to oxidation-state changes without the energy-referencing and fitting details.

The N1 range is 402.0--402.2\,eV, whereas N2 and N3 span 399.1--399.8 and 397.5--397.9\,eV, respectively. At $+0.30$ and $+0.40$\,V, N2 occurs at 399.3 and 399.1\,eV, and N3 at 397.7 and 397.5\,eV. These changes should be read together with the component widths and source attribution limits, rather than treated as independently established new N species. C1 is 288.8--288.9\,eV, C2 is 287.6\,eV in every listed analysis, C3 is 286.1--286.2\,eV, and C4 is 284.6--284.8\,eV. The recurrence of identical rounded values does not establish identical chemical environments or prove that these parameters were unconstrained during fitting.

O\,1s lies at 532.3--532.7\,eV. The K\,2p entry is 293.4\,eV through $+0.20$\,V, followed by 294.0 and 293.8\,eV at the two higher potentials; it is retained as a K\,2p channel rather than assigned a more specific spin--orbit label absent from the export. Reported F\,1s and Ca\,2p$_{3/2}$ energies span 688.6--689.0 and 346.8--347.5\,eV, respectively. The Au entry is 83.9--84.0\,eV where reported, with its source orbital label retained pending verification. Blank N and C total fields are expected because an elemental sum has no unique peak energy.

No uncertainty or calibration accuracy is supplied for these binding-energy entries. The 0.1\,eV reporting increment therefore should not be used as an error bar or as proof that differences of a few tenths of an eV are significant. Full spectra and energy-scale referencing are required before assigning these small shifts to a particular surface reaction or structural transition.

\begin{table}[htbp]
\centering
\caption{Binding energies (eV) for the reported channels and fitted components in the archived export.}
\label{tab:S3}
\scriptsize
\setlength{\tabcolsep}{3.5pt}
\begin{tabular}{@{}llrrrrrrrrrrrrrrr@{}}
\toprule
 & $E_{\mathrm{dep}}$ & Fe & F$^{a}$ & O & \multicolumn{4}{c}{N\,1s} & Ca$^{a}$ & K & \multicolumn{5}{c}{C\,1s} & Au$^{b}$ \\
\cmidrule(lr){6-9}\cmidrule(lr){12-16}
ID & (V vs.\ SCE) & 2p$_{3/2}$ & 1s & 1s & N1 & N2 & N3 & Total & 2p$_{3/2}$ & 2p & C1 & C2 & C3 & C4 & Total & 4f$_{5/2}$ \\
\midrule
10:1 & $-0.20$ & 708.7 & --- & 532.4 & 402.1 & 399.7 & 397.9 & --- & 346.8 & 293.4 & 288.8 & 287.6 & 286.2 & 284.8 & --- & 83.9 \\
11:1 & $-0.20$ & 708.7 & --- & 532.5 & 402.1 & 399.7 & 397.9 & --- & 346.9 & 293.4 & 288.8 & 287.6 & 286.2 & 284.8 & --- & 83.9 \\
12:1 & $-0.20$ & 708.7 & --- & 532.4 & 402.1 & 399.7 & 397.9 & --- & 346.9 & 293.4 & 288.8 & 287.6 & 286.1 & 284.8 & --- & 83.9 \\
13:1 & $-0.20$ & 708.7 & --- & 532.5 & 402.1 & 399.8 & 397.9 & --- & 346.9 & 293.4 & 288.8 & 287.6 & 286.2 & 284.8 & --- & 84.0 \\
1:1 & $-0.10$ & 708.7 & 689.0 & 532.4 & 402.1 & 399.7 & 397.9 & --- & 346.8 & 293.4 & 288.8 & 287.6 & 286.2 & 284.8 & --- & 83.9 \\
2:1 & $-0.10$ & 708.7 & 688.7 & 532.4 & 402.1 & 399.7 & 397.9 & --- & 346.9 & 293.4 & 288.8 & 287.6 & 286.2 & 284.8 & --- & 83.9 \\
3:1 & $-0.10$ & 708.7 & 688.9 & 532.3 & 402.1 & 399.7 & 397.9 & --- & 346.9 & 293.4 & 288.8 & 287.6 & 286.2 & 284.8 & --- & 83.9 \\
4:1 & $-0.10$ & 708.7 & 688.7 & 532.3 & 402.1 & 399.7 & 397.9 & --- & 347.0 & 293.4 & 288.8 & 287.6 & 286.2 & 284.8 & --- & 83.9 \\
5:1 & $0.00$ & 708.7 & 688.6 & 532.4 & 402.1 & 399.7 & 397.9 & --- & 346.9 & 293.4 & 288.8 & 287.6 & 286.2 & 284.8 & --- & 83.9 \\
6:1 & $+0.10$ & 708.7 & 688.9 & 532.3 & 402.1 & 399.7 & 397.9 & --- & 346.9 & 293.4 & 288.8 & 287.6 & 286.2 & 284.8 & --- & --- \\
7:1 & $+0.20$ & 708.7 & 689.0 & 532.4 & 402.1 & 399.7 & 397.9 & --- & 346.9 & 293.4 & 288.8 & 287.6 & 286.2 & 284.8 & --- & --- \\
8:1 & $+0.30$ & 708.5 & 688.8 & 532.6 & 402.0 & 399.3 & 397.7 & --- & 347.5 & 294.0 & 288.8 & 287.6 & 286.1 & 284.7 & --- & --- \\
9:1 & $+0.40$ & 708.3 & --- & 532.7 & 402.2 & 399.1 & 397.5 & --- & --- & 293.8 & 288.9 & 287.6 & 286.1 & 284.6 & --- & --- \\
\bottomrule
\end{tabular}

\vspace{4pt}
\begin{minipage}{\linewidth}
\footnotesize Elemental N and C totals have no single binding energy; their dashes reproduce blank fields in the source. All archived sample names begin with ``PB standard PSD'' and specify 10\,mC deposition charge. At $-0.10$\,V, IDs 1:1, 2:1, 3:1 and 4:1 carry no suffix, ``bis'', ``ter'' and ``quater'', respectively. At $-0.20$\,V, IDs 10:1, 11:1, 12:1 and 13:1 carry no suffix, ``ter'', ``bis'' and ``quater'', respectively. IDs 5:1--9:1 correspond to 0.00 through $+0.40$\,V with no suffix. These suffixes are source identifiers rather than a validated description of independent film replication.

\footnotesize $^{a}$\,F and Ca are identified as survey estimates in the source. A dash denotes a blank source entry, not zero or a demonstrated absence of the element. $^{b}$\,The Au orbital label is reproduced as 4f$_{5/2}$ from the export; the label requires verification against the original Au doublet. N3 and the C-component names are source labels, as explained in Table~\ref{tab:S1}.
\end{minipage}
\end{table}

\section{Peak widths: full width at half maximum}

Table~\ref{tab:S4} reports FWHM values in eV for every channel and fitted component supplied in the export. FWHM is the energy width of a peak at half its maximum height. It is a parameter of the spectral line shape and must be distinguished from the uncertainty in its binding energy, the standard deviation of an atomic ratio, and the energy resolution of the instrument. Measured or fitted widths can include instrumental and lifetime broadening, a distribution of chemical environments, unresolved spectral structure, and effects of the adopted fitting model.\cite{Major2020}

The Fe channel has a FWHM of 1.2\,eV from $-0.20$ to $+0.20$\,V and narrows to 1.0 and 1.1\,eV at $+0.30$ and $+0.40$\,V. These widths belong to the single exported Fe channel and are not a substitute for the widths of every Fe(II), Fe(III), satellite, or multiplet contribution in the separate high-resolution fit. The O\,1s widths are 2.3--2.7\,eV, broader than the N3, K and C entries in this export. A broad O signal may contain overlapping environments, but the width alone cannot distinguish water, hydroxyl groups, and other oxygen-containing species.

N1 and N2 each span 1.1--2.0\,eV across the listed analyses. Their widths reach 1.8\,eV at $+0.30$\,V and 2.0\,eV at $+0.40$\,V, while N3 remains between 0.9 and 1.0\,eV. The simultaneous broadening of N1 and N2 describes the fitted model at high potential; it does not independently validate the chemical attribution or prove formation of a separate phase. The absence of recorded fit constraints prevents deciding whether their equal widths were constrained or obtained independently.

All four C components have widths of 1.1\,eV in the listed analyses through $+0.30$\,V and 1.2\,eV at $+0.40$\,V. Their repeated equality could reflect a shared fitting constraint, but that constraint is not documented and is not assumed here. K spans 0.9--1.1\,eV, F 2.4--2.5\,eV, Ca 1.4--2.3\,eV, and Au is reported as 0.7\,eV wherever a width is supplied. Values for F and Ca remain associated with the source survey estimates and should not be interpreted as verified high-resolution fit results.

The widths provide useful checks when reading the binding energies and component concentrations together. They do not measure crystallite size, film uniformity, purity, or electrochemical performance. A chemically defensible interpretation requires the measured spectra, summed fit, background, individual components, residuals, and the constraints used in fitting.\cite{Major2020} None of the missing fit information has been reconstructed from the tabulated numbers.

\begin{table}[htbp]
\centering
\caption{Full widths at half maximum (eV) of the reported channels and fitted components in the archived export.}
\label{tab:S4}
\scriptsize
\setlength{\tabcolsep}{3.5pt}
\begin{tabular}{@{}llrrrrrrrrrrrrrrr@{}}
\toprule
 & $E_{\mathrm{dep}}$ & Fe & F$^{a}$ & O & \multicolumn{4}{c}{N\,1s} & Ca$^{a}$ & K & \multicolumn{5}{c}{C\,1s} & Au$^{b}$ \\
\cmidrule(lr){6-9}\cmidrule(lr){12-16}
ID & (V vs.\ SCE) & 2p$_{3/2}$ & 1s & 1s & N1 & N2 & N3 & Total & 2p$_{3/2}$ & 2p & C1 & C2 & C3 & C4 & Total & 4f$_{5/2}$ \\
\midrule
10:1 & $-0.20$ & 1.2 & --- & 2.5 & 1.2 & 1.2 & 0.9 & --- & 1.8 & 1.0 & 1.1 & 1.1 & 1.1 & 1.1 & --- & 0.7 \\
11:1 & $-0.20$ & 1.2 & --- & 2.5 & 1.2 & 1.2 & 0.9 & --- & 1.8 & 1.0 & 1.1 & 1.1 & 1.1 & 1.1 & --- & 0.7 \\
12:1 & $-0.20$ & 1.2 & --- & 2.6 & 1.2 & 1.2 & 1.0 & --- & 1.8 & 1.1 & 1.1 & 1.1 & 1.1 & 1.1 & --- & 0.7 \\
13:1 & $-0.20$ & 1.2 & --- & 2.5 & 1.2 & 1.2 & 0.9 & --- & 1.4 & 1.0 & 1.1 & 1.1 & 1.1 & 1.1 & --- & 0.7 \\
1:1 & $-0.10$ & 1.2 & 2.4 & 2.5 & 1.2 & 1.2 & 1.0 & --- & 1.7 & 1.0 & 1.1 & 1.1 & 1.1 & 1.1 & --- & 0.7 \\
2:1 & $-0.10$ & 1.2 & 2.5 & 2.7 & 1.2 & 1.2 & 1.0 & --- & 1.8 & 1.1 & 1.1 & 1.1 & 1.1 & 1.1 & --- & 0.7 \\
3:1 & $-0.10$ & 1.2 & 2.5 & 2.6 & 1.1 & 1.1 & 1.0 & --- & 1.9 & 1.1 & 1.1 & 1.1 & 1.1 & 1.1 & --- & 0.7 \\
4:1 & $-0.10$ & 1.2 & 2.4 & 2.5 & 1.3 & 1.3 & 0.9 & --- & 1.9 & 1.0 & 1.1 & 1.1 & 1.1 & 1.1 & --- & 0.7 \\
5:1 & $0.00$ & 1.2 & 2.4 & 2.7 & 1.2 & 1.2 & 1.0 & --- & 2.1 & 1.0 & 1.1 & 1.1 & 1.1 & 1.1 & --- & 0.7 \\
6:1 & $+0.10$ & 1.2 & 2.5 & 2.6 & 1.2 & 1.2 & 1.0 & --- & 1.9 & 1.0 & 1.1 & 1.1 & 1.1 & 1.1 & --- & --- \\
7:1 & $+0.20$ & 1.2 & 2.4 & 2.6 & 1.1 & 1.1 & 0.9 & --- & 2.0 & 1.0 & 1.1 & 1.1 & 1.1 & 1.1 & --- & --- \\
8:1 & $+0.30$ & 1.0 & 2.5 & 2.7 & 1.8 & 1.8 & 0.9 & --- & 2.3 & 1.0 & 1.1 & 1.1 & 1.1 & 1.1 & --- & --- \\
9:1 & $+0.40$ & 1.1 & --- & 2.3 & 2.0 & 2.0 & 1.0 & --- & --- & 0.9 & 1.2 & 1.2 & 1.2 & 1.2 & --- & --- \\
\bottomrule
\end{tabular}

\vspace{4pt}
\begin{minipage}{\linewidth}
\footnotesize Elemental N and C totals have no single FWHM. Reported widths are fit parameters, not uncertainties in peak energy or independently measured instrumental resolutions. The concentration, binding-energy and FWHM tables reproduce all 13 archived analyses, 13 peak/component channels and two N/C elemental-total columns. The three CSV files use the same analysis IDs and ordering; no new fitting, smoothing, peak reassignment or synthetic replicate generation was performed.

\footnotesize $^{a}$\,F and Ca are identified as survey estimates in the source. A dash denotes a blank source entry, not zero or a demonstrated absence of the element.
$^{b}$\,The Au orbital label is reproduced as 4f$_{5/2}$ from the export; the label requires verification against the original Au doublet. N3 and the C-component names are source labels, as explained in Table~\ref{tab:S1}.
\end{minipage}
\end{table}

\renewcommand{\refname}{Supplementary references}